\documentclass[12pt,prd,a4paper,preprint,eqsecnum]{revtex4}
  
\usepackage{amsfonts}
\usepackage{amssymb}
\usepackage{graphicx}
\usepackage{pstricks} 
\usepackage{axodraw}

\newcommand{\be}{\begin{equation}}
\newcommand{\bea}{\begin{eqnarray}}
\newcommand{\ee}{\end{equation}}
\newcommand{\eea}{\end{eqnarray}}
\newcommand{\bpi}{\begin{picture}}
\newcommand{\bce}{\begin{center}}
\newcommand{\epi}{\end{picture}}
\newcommand{\ece}{\end{center}}

\begin{document}

\begin{flushright}
ECT*-06-20\\
FTUV-2006-1127
\end{flushright}


\title{Pinch Technique for Schwinger-Dyson Equations}

\author{Daniele Binosi}
\email{binosi@ect.it}
\affiliation{ECT* Villa Tambosi, Strada delle Tabarelle 286,
I-38050 Villazzano (Trento), Italy}

\author{Joannis Papavassiliou}
\email{joannis.papavassiliou@uv.es}
\affiliation{Departamento de F\'\i sica Te\'orica and IFIC
Centro Mixto, Universidad de Valencia (Fundaci\'on General), E-46100, Burjassot,
Valencia, Spain}


\begin{abstract}

In  the context  of scalar  QED we
derive the  pinch technique  self-energies and vertices  directly from
the  Schwinger-Dyson  equations.   After  reviewing  the  perturbative
construction,  we discuss in  detail the  general methodology  and the
basic field-theoretic ingredients necessary for the completion of this
task.   The construction  requires the  simultaneous treatment  of the
equations  governing  the   scalar  self-energy  and  the  fundamental
interaction  vertices.   The  resulting non-trivial  rearrangement  of
terms  generates  dynamically the  Schwinger-Dyson  equations for  the
corresponding Green's  functions of the background  field method.  The
proof  relies on the  extensive use  of the  all-order Ward-identities
satisfied   by  the   full  vertices   of  the   theory  and   by  the
one-particle-irreducible  kernels  appearing  in  the  usual  skeleton
expansion.   The  Ward  identities  for these  latter  quantities  are
derived formally,  and several subtleties related to  the structure of
the multiparticle kernels are addressed.  The general strategy for
the generalization of  the method in a non-Abelian  context is briefly
outlined, and some of the technical difficulties are discussed.

\end{abstract}

\maketitle


\section{\label{intro} Introduction}

The most widely  used framework for studying in  the continuum various
dynamical  questions  that  lie  beyond perturbation  theory  are  the
Schwinger-Dyson  equations (SDE)~\cite{Dyson:1949ha,Schwinger:1951ex}.
This infinite system of  coupled non-linear integral equations for all
Green's  functions of  the theory  is inherently  non-perturbative, and
captures the  full content of  the quantum equations of  motion.  Even
though  these  equations  are   derived  by  an  expansion  about  the
free-field  vacuum,  they finally  make  no  reference  to it,  or  to
perturbation theory,  and can be  used to address problems  related to
chiral  symmetry  breaking, dynamical  mass  generation, formation  of
bound      states,     and     other      non-perturbative     effects
\cite{Cornwall:1974vz,Marciano:su}.   Since  this  system involves  an
infinite hierarchy  of equations, in practice one  is severely limited
in their use, and the  need for a self-consistent truncation scheme is
evident.  Devising such  a scheme, however, is far  from trivial, even
in  the  case   of  toy  models~\cite{Korpa:1990jp,Sauli:2001mb},  and
becomes far more challenging when dealing with gauge theories.  One of
the central problems  in this latter context stems  from the fact that
the  SDEs are  built out  of unphysical  Green's functions;  thus, the
extraction  of  reliable  physical  information depends  crucially  on
delicate all-order cancellations, which may be inadvertently distorted
in the  process of the truncation.   In QED the issues  related to the
truncation         of         the         SDE         are         very
delicate~\cite{Kondo:1988md,Curtis:1990zs,Curtis:1990zr,Bashir:1997qt,Sauli:2002tk},
but  the  level of  complexity  increases  further  when dealing  with
non-Abelian   gauge   theories~\cite{Mandelstam:1979xd},   where   the
ghost-infested  Slavnov-Taylor identities  (STI) \cite{Slavnov:1972fg}
make time-honored methods, such as the ``gauge
technique''~\cite{Salam:1963sa}, much more difficult to implement.

The   truncation   scheme   based   on  the   pinch   technique   (PT)
~\cite{Cornwall:1982zr}   attempts  to   address  this
problem at  its root, introducing a drastic modification  already at
the  level  of  the building  blocks  of  the  SD series,  namely  the
off-shell  Green's functions  themselves.   The PT  is a  well-defined
algorithm  that  exploits  systematically  the symmetries  built  into
physical observables, such as  $S$-matrix elements or Wilson loops, in
order to construct new,  effective Green's functions, endowed with very
special  properties, generally  associated with  physical observables.
The basic observation, which essentially defines the PT, is that there
exists  a  fundamental  cancellation  between sets  of  diagrams  with
different kinematic  properties, such as  self-energies, vertices, and
boxes.  This  cancellation is driven  by the underlying  BRST symmetry
\cite{Becchi:1976nq}, and  is triggered when a  very particular subset
of the longitudinal momenta circulating inside vertex and box diagrams
generate   out  of   them   (by  ``pinching''  internal   lines)
propagator-like  terms.   The latter  are  reassigned to  conventional
self-energy  graphs,  in order  to  give  rise  to the  aforementioned
effective  Green's   functions.   These  new   Green's  functions  are
independent of the  gauge-fixing 
parameter~\cite{Cornwall:1982zr,Cornwall:1989gv,Papavassiliou:1989zd,Papavassiliou:1994pr,Binosi:2001hy},
satisfy ghost-free,
QED-like  Ward identities  (WI) instead  
of the  complicated STI~\cite{Cornwall:1989gv,Papavassiliou:1989zd},
display only  physical  thresholds~\cite{Papavassiliou:1995fq,Papavassiliou:1996fn},  
have correct analyticity properties~\cite{Papavassiliou:1996zn}, 
and are  well-behaved  
at high energies~\cite{Papavassiliou:1997fn}. In addition, as has been shown recently~\cite{Binger:2006sj},
the form factors of the one-loop PT three-gluon vertex~\cite{Cornwall:1989gv}
satisfy relations characteristic of supersymmetric scattering amplitudes.
For some recent application of the PT in  
non-commutative theories, see~\cite{Caporaso:2005xf,Caporaso:2006ij}.
Returning to the SDEs,   
the final upshot  of the PT program is  to 
trade  the conventional  SD series  for another,  written in  terms of
these  new Green's  functions, and  then truncate  it, keeping
only a few  terms in a ``dressed-loop'' expansion,  maintaining at the
same time exact gauge-invariance~\cite{Cornwall:1982zr} .

Due to various theoretical advances in recent years, the PT
has been  generalized to all  orders in perturbation theory,  both for
QCD \cite{Binosi:2002ft}  and the  electroweak sector  
of the  Standard Model \cite{Binosi:2004qe}.   
Of central importance in  this context is the  connection between the  PT and the
Background Field  Method (BFM)~\cite{Dewitt:ub}.  The latter is  a special gauge-fixing
procedure  that preserves the  symmetry of  the action  under ordinary
gauge transformations with respect to the background (classical) gauge
field   $\widehat{A}^a_{\mu}$,   while   the  quantum   gauge   fields,
$A^a_{\mu}$, appearing  in the loops, transform  homogeneously under the
gauge  group~\cite{Weinberg:kr}.  As  a   result,  the  background
$n$-point  functions satisfy QED-like  all-order WIs.   
Note that the BFM gives rise to special Feynman rules and a 
characteristic ghost sector.
The connection
between PT and  BFM~\cite{Denner:1994nn}, 
known to persist to all  orders~\cite{Binosi:2002ft}, 
affirms that the
(gauge-independent) PT effective $n$-point functions coincide with the
(gauge-dependent) BFM $n$-point functions provided that the latter are
computed in the Feynman gauge.

Despite this progress, however,  the truncation program outlined above
is still incomplete.  In fact,  the direct implementation of the PT at
the level of  the SDE is an entirely  unexplored question.  Of course,
PT-inspired SDEs have been treated  in the 
literature~\cite{Cornwall:1982zr,Cornwall:1989gv,Papavassiliou:1991hx,Mavromatos:1999jf}, 
but rather than
derived they have been  postulated heuristically, based on perturbative
diagrammatic analysis.  In  the most recent work in  this direction~\cite{Aguilar:2006gr}, 
the SDE for the PT gluon propagator was formulated  
directly in  the  BFM ~\cite{Sohn:1985em}; there, 
the connection between  the PT and  the BFM has been extrapolated without proof 
from perturbation theory to the SDE. 
In the present paper we take the
first step toward  the full implementation of the PT  at the level of
SDE.  Specifically, we will carry out the PT rearrangement for the SDE
of {\it  scalar QED}.  This Abelian  model captures a  plethora of the
relevant  conceptual issues, without  the additional  complications of
non-Abelian  theories,  and serves  as  an  excellent  toy theory  for
gaining valuable insight on the problem.
   
The  application of  the PT  in  an Abelian  context might  seem as  a
trivial exercise  at first,  but this is  certainly not the  case: the
self-energy  of  the charged  scalar  undergoes non-trivial  pinching,
displaying  a  great  deal  of  the  characteristics  known  from  the
non-Abelian  studies.  The reason  for this  may be  traced back  to a
simple  fact,  namely  the  momentum  dependence of  the  bare  vertex
describing the interaction between the scalars and the photon. This is
exactly analogous  to what happens with the  three-gluon vertex, which
is the  central object when carrying  out the PT  construction in QCD.
According to the  standard PT methodology, from  the scalar-photon vertex
one isolates  its ``pinching'' part, {\it i.e.},  the  combination of momenta
that trigger  the standard elementary WI when  contracted with another
such vertex. The terms generated from this WI are to be reassigned and
interpreted  following exactly  the standard  PT  philosophy, arriving
eventually at a new modified scalar self-energy.

The main result of this  article is the following.  The application of
the  PT at  the  level of  the SDEs  obtained  in the  context of  the
covariant  gauges for  the conventional  Green's  functions, generates
{\it  dynamically} the  corresponding SDEs  governing the  BFM Green's
functions.   Operationally this  is accomplished  following  the basic
rules established  from the  perturbative analysis, sublemented  by an
additional crucial step. 
Specifically, when dealing with the SDE
for the  scalar self-energy, one  must pinch {\it  simultaneously} the
SDE's governing the full vertices.   It is only then that the ensuing,
highly non-trivial  rearrangements conspire to  generate {\it exactly}
the  terms  responsible  for  the   conversion  of  the  SDE  for  the
conventional   scalar  self-energy   into  the   SDE  for   the  BFM-PT
self-energy.

Instrumental for  the implementation  of the procedure  outlined above
are three ingredients: ({\it i}) the all-order WI relating the divergence of
the  full photon-scalar  vertex with  the scalar  self-energy;  ({\it ii}) the
all-order   WI  satisfied  by   the  one-particle   irreducible  (1PI)
multi-particle  kernels appearing  in  the skeleton  expansion of  the
SDE's  for the  relevant full  vertices;  ({\it iii}) a  set of  non-trivial
identities~\cite{Gambino:1999ai}, relating the BFM $n$-point functions
to the corresponding conventional $n$-point functions in the covariant
renormalizable gauges,  to all  orders in perturbation  theory.  These
identities,  to  be   referred  to  as  Background-Quantum  identities
(BQIs),   furnish   a  concrete   field-theoretic
identification  of the  terms that  are removed  during  the pinching
procedure from the conventional Green's functions, in order to generate
their BFM counterparts~\cite{Binosi:2002ez}.

The  article is organized  as follows.   In Section \ref{sec2} we  review some
general  characteristics of  scalar QED  and its  quantization  in the
conventional covariant gauges, present the all-order WI's for
the  two fundamental vertices,  and derive  formally the  relevant BQIs
relating the  scalar self-energy and vertices in  the covariant gauges
with the corresponding quantities in the BFM. In section \ref{ptpert} we present
a brief review of the PT applied  to the case of scalar QED, and carry
out explicitly  the construction of  the PT scalar self-energy,  at one
and two loops.  In addition  to setting up the notation and describing
the general philosophy, this presentation  serves as a warm-up for the
generalization of  the method at the  level of the  SDEs.  Therefore, we
pay particular attention to  the general patterns appearing already at
two loops, with special emphasis on how to reorganize various diagrams
in  order to  identify  the larger  structures  (Green's functions  or
kernels) on which the pinching momenta  act.  In Section \ref{sdq} we give a
qualitative  discussion of  the general  strategy we  will  follow when
pinching the  SDEs, comment on the technical  subtleties, and determine
the necessary ingredients for  the implementation of this program.  In
Section \ref{ck}  we derive in  detail the all-order  WI for two of  the 1PI
kernels appearing in the SDE's.  Section \ref{sde} contains the main thrust of
our paper: using the machinery  developed in the previous section, the
PT construction is carried out explicitly for the SDE's of scalar QED.
In   Section \ref{Conc}  we   present  our   conclusions  and   discuss  the
generalization of  this work to  a non-Abelian context.   Finally, the
relevant Feynman rules are presented in an Appendix.

\section{\label{sec2} Scalar QED and its identities}

In this section we present the Lagrangian of scalar QED and 
the procedure of its gauge-fixing, in the context of both conventional 
renormalizable gauges and BFM. We derive the all-order WI's satisfied by 
the fundamental vertices of the theory, and  
useful identities (BQI) relating the Green's functions of the theory
in the two aforementioned gauge-fixing schemes.

\subsection{Lagrangian and gauge fixing}

We will concentrate on scalar QED, which describes a complex 
scalar field $\phi$ interacting with the electromagnetic field $A_\mu$. The Lagrangian density is 
\begin{equation}
{\cal L} = {\cal L}_{\mathrm I} + {\cal L}_{\mathrm{GF}} + {\cal L}_{\mathrm{FPG}},
\label{SQED_lag}
\end{equation}
with ${\cal L}_{\mathrm I}$ the gauge invariant $U(1)$ Lagrangian,
\begin{equation}
{\cal L}_{\mathrm I} = -\frac14 F^{\mu\nu}F_{\mu\nu}+\left({\cal D}^\mu\phi\right)^\dagger\left({\cal D}_\mu\phi\right)-m^2\phi^\dagger\phi+\frac\lambda4(\phi^\dagger\phi)^2 ,
\label{Linv}
\end{equation}
where the field strength is
\begin{equation}
F_{\mu\nu}=\partial_\mu A_\nu-\partial_\nu A_\mu,
\end{equation}
and the covariant derivative is defined as
\begin{equation}
{\cal D}_\mu=\partial_\mu-igY_\phi A_\mu,
\end{equation}
with $g$ the coupling constant and $Y_\phi$ the scalar field hypercharge ($Y_\phi=1$).
In general, the (covariant) gauge fixing term and the Faddeev-Popov ghost can be written as
\begin{eqnarray}
{\cal L}_{\mathrm{GF}}&=&\frac\xi2B^2+B{\cal F},\\
{\cal L}_{\mathrm{FPG}} &=&-\bar c s{\cal F}.
\end{eqnarray}
In the formulas above, ${\cal F}$ is the gauge fixing function, $B$ is an auxiliary, 
non-dynamical field that can be eliminated through its (trivial) equation of motion, $c$ ($\bar c$) is the ghost (anti-ghost) field,
while $s$ is the BRST operator, with
\begin{eqnarray}
sA_\mu=\partial_\mu c &\qquad& s\phi=ig c\phi, \nonumber \\
s\phi^\dagger=-ig c\phi^\dagger &\qquad& sc=0, \nonumber \\
s\bar c=B &\qquad& sB=0.
\end{eqnarray}
In view of the equivalence between the PT Green's functions and the BFM ones at $\xi_{\mathrm Q}=1$ we will consider the
following two gauge fixing procedures (the corresponding Feynman rules relevant for our calculation are given in the Appendix). 

\begin{enumerate}

\item In the usual $R_\xi$ gauges, one chooses ${\cal F}=\partial^\mu A_\mu$, to get
\begin{eqnarray}
{\cal L}_{\mathrm{GF}}&=&-\frac1{2\xi}(\partial^\mu A_\mu)^2,\\
{\cal L}_{\mathrm{FPG}} &=&-\bar c \partial^2c.
\end{eqnarray}
In this gauge the ghosts are, of course, decoupled and play no dynamical role.
 
\item In the case of the BFM, one splits the scalar field into a 
background part, $\widehat\phi$, and its quantum part, $\phi$. 
Notice that the BRST variation of the background field will be zero, 
but the latter will enter in the variation of the quantum one, {\it i.e.},
\begin{eqnarray}
s\phi=ig c(\widehat\phi+\phi) &\qquad& s\phi^\dagger=-ig c(\widehat\phi^\dagger+\phi^\dagger)\\
s\widehat\phi=0 &\qquad& s\widehat\phi^\dagger=0.
\end{eqnarray}
In this case the gauge fixing function is 
\begin{equation}
{\cal F}=\partial^\mu A_\mu -ig\xi(\widehat\phi^\dagger\phi-\phi^\dagger\widehat\phi),
\label{BFMgff}
\end{equation} 
which gives in turn 
\begin{eqnarray}
{\cal L}_{\mathrm{GF}}&=&-\frac1{2\xi}(\partial^\mu A_\mu)^2+
ig\partial^\mu A_\mu(\widehat\phi^\dagger\phi-\phi^\dagger\widehat\phi)\nonumber \\ 
&+& \frac{g^2}2\xi\left[(\widehat\phi^\dagger\phi)^2+(\phi^\dagger\widehat\phi)^2  
-2\widehat\phi^\dagger\phi\phi^\dagger\widehat\phi\right], \\
{\cal L}_{\mathrm{FPG}}&=&-\bar c\partial^2c-g^2\xi \bar c\left(\widehat\phi^\dagger c\phi
+2\widehat\phi^\dagger c \widehat\phi+\phi^\dagger c\widehat\phi\right).
\label{L-BFM}
\end{eqnarray}
Therefore, in the BFM ghosts are not decoupled. 
In order to obtain the full set of Feynman rules in this gauge, 
one needs also to consider the extra terms 
coming from the background-quantum splitting carried out inside the 
gauge invariant Lagrangian, {\it i.e.}, 
${\cal L}_{\mathrm I}(\phi,\phi^\dagger)\to{\cal L}_{\mathrm I}(\widehat\phi+\phi,\widehat\phi^\dagger+\phi^\dagger)$, see again the Appendix.
\end{enumerate}

\subsection{Some fundamental identities}

In this subsection we review briefly the Batalin-Vilkovisky formalism~\cite{Batalin:pb}, which allows one to get 
simultaneously both the WIs as well as the BQIs of our theory.

Let us  then start by introducing  for each field  $\Phi$ appearing in
the  theory  the  corresponding  anti-field, to be denoted by $\Phi^*$. 
The anti-field $\Phi^*$
has opposite  statistics   with  respect   to  $\Phi$; its ghost number, ${\rm gh}(\Phi^*)$,
is related to the  ghost  number ${\rm gh}(\Phi)$ of the field $\Phi$ by 
${\rm gh}(\Phi^*)=-1- {\rm gh}(\Phi)$.
The ghost numbers of the various fields and anti-fields are summarized
in table I.
\begin{table}
	\centering
		\begin{tabular}{|c|c|c|c|c||c|c|c|c|c|}
		\hline
		\mbox{}$\Phi$ & $A_\mu$ & $\phi$ & $c$ & $\bar c$ & $A^*_\mu$ & $\phi^*$ & $c^*$ & $\bar c^*$\\
		\hline
		${\rm gh}(\Phi)$ & 0 & 0 & 1 & -1 & -1 & -1 & 0 & -2\\
		\hline
		\end{tabular}
		\caption{The ghost numbers of the fields and antifields in scalar QED.}
\end{table}
Next, we add to the original gauge invariant Lagrangian a 
term coupling the anti-fields with the BRST variation of the corresponding fields, to get
\begin{eqnarray}
{\cal L}_{\mathrm{BV}}&=&{\cal L}_{\mathrm I}+{\cal L}_{\mathrm{BRST}} \nonumber \\
&=& {\cal L}_{\mathrm I} +\sum_{\Phi}\Phi^*s\Phi.
\end{eqnarray}
As a consequence of the BRST invariance of the action and the nilpotency of the 
BRST operator, the action $\Gamma^{(0)}[\Phi,\Phi^*]$ constructed from ${\cal L}_{\mathrm{BV}}$, 
will satisfy the master equation
\begin{equation}
\int\!\!d^4x\sum_\Phi\frac{\delta\Gamma^{(0)}}{\delta\Phi^*}\frac{\delta\Gamma^{(0)}}{\delta\Phi}=0.
\label{master_eq}
\end{equation}
Since the anti-fields are external  sources, we must constrain them to
suitable values before we  use the action $\Gamma^{(0)}$ in $S$-matrix
elements  calculations \cite{Weinberg:kr}. To  that end, we  introduce an  arbitrary
fermionic  functional $\Psi[\Phi]$ (also referred to as ``gauge  fixing
fermion'', for reasons that will  become clear shortly), such
that
\begin{equation}
\Phi^*=\frac{\delta\Psi[\Phi]}{\delta\Phi}.
\end{equation}
Then the action becomes
\begin{eqnarray}
\Gamma^{(0)}[\phi,\delta\Psi/\delta\Phi]&=&\Gamma^{(0)}[\Phi]+(s\Phi)\frac{\delta\Psi[\Phi]}{\delta\Phi}\nonumber \\
&=& \Gamma^{(0)}[\Phi]+s\Psi[\Phi],
\end{eqnarray}
and choosing the functional $\Psi$ to satisfy the relation
\begin{equation}
s\Psi=\int\!d^4x\left({\cal L}_{\mathrm{GF}}+{\cal L}_{\mathrm{FPG}}\right),
\end{equation}  
we see that the action $\Gamma^{(0)}$ (obtained from ${\cal L}_{\mathrm{BV}}$) 
is equivalent to the gauge fixed action obtained from the original Lagrangian ${\cal L}$ of Eq.(\ref{SQED_lag}). 
Furthermore, the antighost anti-field $\bar c^*$ and the multiplier $B$ enter bi-linearly in the action, 
and one can write
\begin{equation}
\Gamma^{(0)}[\Phi,\Phi^*]=\Gamma^{(0)}_{\mathrm{min}}[A_\mu,A^*_\mu,\phi,\phi^*,\phi^\dagger,\phi^{*\dagger},c,c^*]-B\bar c^*,
\end{equation}
with $\Gamma^{(0)}_{\mathrm{min}}$ satisfying 
the master equation (\ref{master_eq}) by itself. 
In what follows we will restrict our considerations to 
the minimal action, dropping the corresponding subscript.

The quantum corrected version of the master equation (\ref{master_eq}) is established in the form of the WI functional
\begin{equation}
{\cal S}(\Gamma)[\Phi,\Phi^*]=\int\!d^4x\sum_\Phi\frac{\delta\Gamma}{\delta\Phi^*}\frac{\delta\Gamma}{\delta\Phi}=0,
\label{STIfunc}
\end{equation}
where $\Gamma[\Phi,\Phi^*]$ is now the effective action. 
The equation above must hold in any theory with a unitary $S$-matrix 
and gauge-independent physical observables, and gives rise to the 
complete set of the all-order WIs, 
via the repeated application of functional differentiations, keeping in mind that: 
({\it i}) ${\cal S}(\Gamma)$ has ghost charge 1; ({\it ii}) functions 
with non-zero ghost charge vanish (since the ghost charge is a conserved quantity); 
({\it iii}) the BRST transformation of the gauge field is proportional to the ghost $sA_\mu=\partial_\mu c$.
Overall, these latter observations imply that in order to extract
non-zero identities from Eq.(\ref{STIfunc}) one needs to differentiate the latter with respect to one ghost filed, or two ghost fields and one anti-field (the only exception to this rule is when differentiating with respect to a ghost anti-field.) In particular, identities involving one or more gauge fields are obtained differentiating Eq.(\ref{STIfunc})
with respect to the set of fields in which one gauge boson has been replaced by the corresponding ghost field.

In the remainder of this section we will adopt for the $n$-point Green's function the notation
\begin{equation}  
\Gamma_{\Phi_1\Phi_2...\Phi_{n-1}\Phi_n}(p_2,...,p_{n-1},p_n) = i^n\left.\frac{\delta^{n}\Gamma}{\delta\Phi_1(p_1)\delta\Phi_2(p_2)\cdots\delta\Phi_n(p_n)}\right|_{\Phi_i=0},
\end{equation} 
with $p_i$ the in-going momentum of the $\Phi_i$ field. 
The momentum for the field $\Phi_1$ follows from momentum conservation, and we will not write it explicitly.

\subsubsection{Ward Identities}

We begin with the STI  
for the photon-scalar-scalar vertex. From our previous discussion 
follows that, for obtaining such an identity, one needs to consider the functional differentiation
\begin{equation}
\left.\frac{\delta^3{\cal S}(\Gamma)}{\delta\phi^\dagger(p_1)\delta\phi(p_2)\delta c(q)}\right|_{\Phi=0}=0 \qquad p_1+p_2+q=0.
\end{equation}
Carrying out the functional differentiation we obtain the equation 
\be
\Gamma_{cA^*_\mu}(q)\Gamma_{A_\mu\phi^\dagger\phi}(p_1,p_2) +
\Gamma_{\phi^\dagger c\phi^*}(q,p_2)\Gamma_{\phi^\dagger\phi}(p_2) 
+ \Gamma_{\phi c\phi^{*\dagger}}(q,p_1)\Gamma_{\phi\phi^\dagger}(p_1)=0
\ee
On the other hand, the Abelian nature of the theory, together with the decoupling of the ghosts in the $R_\xi$ gauges, enforce the validity 
of the following (all order) equations (see also the Appendix)
\begin{eqnarray}
& & \Gamma_{cA^*_\mu}(q)=\Gamma^{(0)}_{cA^*_\mu}(q)=-q^\mu,\nonumber \\
& & \Gamma_{\phi^\dagger c\phi^*}(q,p_2) = \Gamma^{(0)}_{\phi^\dagger c\phi^*}(q,p_2)=g, \nonumber \\
& & \Gamma_{\phi c\phi^{*\dagger}}(q,p_1) = \Gamma^{(0)}_{\phi c\phi^{*\dagger}}(q,p_1)=-g,
\label{gdec}
\end{eqnarray}
which furnish  the fundamental WI
\begin{equation}
q^\mu\Gamma_{A_\mu\phi^\dagger\phi}(p_1,p_2)=g\left[\Gamma_{\phi\phi^\dagger}(p_2)-\Gamma_{\phi\phi^\dagger}(p_1)\right].
\label{WI3-l}
\end{equation}
Introducing the short-hand notation $\Gamma_{A_\mu\phi^\dagger\phi}\equiv\Gamma_\mu$ and $\Gamma_{\phi\phi^\dagger}\equiv S^{-1}$, 
we finally get 
\begin{equation}
q^\mu\Gamma_\mu(p_1,p_2)=g\left[S^{-1}(p_2)-S^{-1}(p_1)\right].
\label{WI3}
\end{equation}

Let us derive next the WI satisfied by the 4-point 
function $\Gamma_{A_\mu A_\nu\phi^\dagger\phi}$. In this case we need to consider the functional differentiation
\begin{equation}
\left.\frac{\delta^4{\cal S}(\Gamma)}{\delta\phi^\dagger(p_1)\delta\phi(p_2)\delta A_\nu(k)\delta c(q)}\right|_{\Phi=0}=0 
\qquad p_1+p_2+k+q=0.
\end{equation}
Carrying out the functional differentiation, 
and considering that, due to the decoupling of the ghost fields, 
one has to all orders
\begin{eqnarray}
\Gamma_{A_\nu cA^*_\mu}=0 &\qquad& \Gamma_{\phi^\dagger\phi cA^*_\mu}=0, \nonumber \\
\Gamma_{\phi^\dagger A_\nu c \phi^*}=0 &\qquad& \Gamma_{\phi A_\nu c\phi^{*\dagger}}=0, \nonumber
\end{eqnarray}
we obtain
\begin{eqnarray}
\Gamma_{cA^*_\mu}(q)\Gamma_{A_\mu A_\nu\phi^\dagger\phi}(k,p_1,p_2)&+&
\Gamma_{\phi^\dagger c\phi^*}(q,k+p_2)\Gamma_{\phi^\dagger  A_\nu\phi}(k,p_2)\nonumber \\
&+&\Gamma_{\phi c\phi^{*\dagger}}(q,k+p_1)\Gamma_{\phi A_\nu\phi^\dagger}(k,p_1)=0.
\end{eqnarray}
Then, using once again the results of Eq.(\ref{gdec}), one gets the final identity
\begin{eqnarray}
q^\mu\Gamma_{A_\mu A_\nu\phi^\dagger\phi}(k,p_1,p_2)&=&
g\left[\Gamma_{\phi^\dagger  A_\nu\phi}(k,p_2)-\Gamma_{\phi A_\nu\phi^\dagger}(k,p_1)\right]\nonumber \\
&=&g\left[\Gamma_{A_\nu\phi^\dagger\phi}(-k-p_2,p_2)-\Gamma_{A_\nu\phi^\dagger\phi}(p_1,-k-p_1)\right],
\label{WI4-l}
\end{eqnarray}
which can be rewritten as
\begin{equation}
q^\mu\Gamma_{\mu\nu}(k,p_1,p_2)=g\left[\Gamma_{\nu}(-k-p_2,p_2)-\Gamma_{\nu}(p_1,-k-p_1)\right],
\label{WI4}
\end{equation}
where we have set $\Gamma_{A_\mu A_\nu\phi^\dagger\phi}\equiv\Gamma_{\mu\nu}$.

\subsubsection{Background quantum identities}

Background  quantum  identities  are identities  that  relate
Green's  functions involving  background fields  to  Green's functions
involving only  quantum ones. Therefore, they are  particularly useful in the PT context,
since they  allow for a direct  comparison between PT  and BFM
Green's functions.

To obtain such identities for scalar QED, we introduce a classical scalar field
$\Omega^\phi$   and  its  complex   conjugate  $\Omega^{\phi^\dagger}$,
carrying  the same  quantum numbers as the scalar,  but with ghost  charge
+1. We then implement the  equation of motion of the background fields
at  the quantum  level,  by  extending the  BRST  symmetry through  the
equations
\begin{eqnarray}
s\widehat\phi=\Omega^\phi &\qquad& s\Omega^\phi=0, \nonumber \\
s\widehat{\phi^\dagger}=\Omega^{\phi^\dagger} &\qquad& s\Omega^{\phi^\dagger}=0.
\label{extBRST}
\end{eqnarray}

The dependence of the Green's function on the background field is then controlled by the modified STI functional
\begin{eqnarray}
{\cal S}'(\Gamma')[\Phi,\Phi^*]&=&{\cal S}(\Gamma')[\Phi,\Phi^*]+\int\!d^4x\left[
\Omega^\phi\left(\frac{\delta\Gamma'}{\delta\widehat\phi}-\frac{\delta\Gamma'}{\delta\phi}\right)+
\Omega^{\phi^\dagger}\left(\frac{\delta\Gamma'}{\delta\widehat\phi^\dagger}-\frac{\delta\Gamma'}{\delta\phi^\dagger}\right)\right]\nonumber \\
&=&\int\!d^4x\left\{
\frac{\delta\Gamma'}{\delta A^*_\mu}\frac{\delta\Gamma'}{\delta A_\mu}
+\frac{\delta\Gamma'}{\delta c^*}\frac{\delta\Gamma'}{\delta c}
+\frac{\delta\Gamma'}{\delta\phi^*}\frac{\delta\Gamma'}{\delta\phi^\dagger}
+\frac{\delta\Gamma'}{\delta\phi^{\dagger*}}\frac{\delta\Gamma'}{\delta\phi}\right.\nonumber \\
&+&\left.\left[
\Omega^\phi\left(\frac{\delta\Gamma'}{\delta\widehat\phi}-\frac{\delta\Gamma'}{\delta\phi}\right)+
\Omega^{\phi^\dagger}\left(\frac{\delta\Gamma'}{\delta\widehat\phi^\dagger}-\frac{\delta\Gamma'}{\delta\phi^\dagger}\right)\right]\right\},
\end{eqnarray}
where  $\Gamma'$ denotes  the  effective action  that  depends on  the
background sources  ($\Gamma\equiv\Gamma'\vert_{\Omega=0}$). 
Differentiation of the  above functional with respect to
background  sources and  background and/or  quantum fields  will then
relate 1PI  functions involving different background/quantum field content.

The first BQI we derive involves two background
scalar fields. One begins by considering the functional differentiations
\begin{eqnarray}
\left.\frac{\delta^2{\cal S'}(\Gamma')}{\delta\Omega^{\phi^\dagger}(p_1)\delta\widehat\phi(q)}\right|_{\Phi=0}=0 &\qquad& q+p_1=0, 
\nonumber \\
\left.\frac{\delta^2{\cal S'}(\Gamma')}{\delta\Omega^{\phi^\dagger}(p_1)\delta\phi(q)}\right|_{\Phi=0}=0 &\qquad& q+p_1=0
\end{eqnarray}
which furnish the intermediate BQIs
\begin{eqnarray}
\Gamma_{\widehat\phi\widehat\phi^\dagger}(q)&=&\Gamma_{\widehat\phi\phi^\dagger}(q)+
\Gamma_{\Omega^{\phi^\dagger}\phi^*}(q)\Gamma_{\widehat\phi\phi^\dagger}(q)+
\Gamma_{\Omega^{\phi^\dagger} A^*_\mu}(q)\Gamma_{\widehat\phi A_\mu}(q),\\
\Gamma_{\widehat\phi\phi^\dagger}(q)&=&\Gamma_{\phi\phi^\dagger}(q)+
\Gamma_{\Omega^{\phi}\phi^{*\dagger}}(q)\Gamma_{\phi\phi^\dagger}(q)+
\Gamma_{\Omega^\phi A^*_\mu}(q)\Gamma_{\phi^\dagger A_\mu}(q).
\label{BQI}
\end{eqnarray}
According to our previous discussion, in the above equation all Green's function involving ghost legs have dropped out (having ghost charge different from zero).
The Abelian nature of the theory enforces to all orders the identity
\begin{equation}
\Gamma_{\Omega^{\phi^\dagger} A^*_\mu}(q) = \Gamma_{\Omega^\phi A^*_\mu}(q)=0,
\end{equation} 
so that the BQI relating the background 1PI two-point function to the quantum one reads
\begin{equation}
\Gamma_{\widehat\phi\widehat\phi^\dagger}(q)=
[1+\Gamma_{\Omega^{\phi^\dagger}\phi^*}(q)]^2\Gamma_{\phi\phi^\dagger}(q). 
\label{BQIfinal-l}
\end{equation}
Introducing the auxiliary function $G\equiv\Gamma_{\Omega^{\phi^\dagger}\phi^*}$, 
and denoting $\Gamma_{\widehat\phi\widehat\phi^\dagger}\equiv \widehat S^{-1}$,
we obtain the BQI in its final form,
\begin{equation}
\widehat S^{-1}(q)=[1+G(q)]^2S^{-1}(q). 
\label{BQIfinal}
\end{equation}
Notice that, in the Abelian case, the 1PI Green's function $G$
has a particularly simple expression, namely
\begin{eqnarray}
\label{aux-function}
i G(q)= \mbox{\hspace{3.2cm}}=-g^2\int\! [dk]\frac 1{k^2} S (k+q).
  \begin{picture}(90,35) (435,-222)
    \SetWidth{0.5}
    \SetColor{Black}
    \Text(290,-232)[]{\small{\Black{$\Omega^{\phi^\dagger}(q)$}}}
    \Text(235,-232)[]{\small{\Black{$\phi^*(q)$}}}
    \SetWidth{0.5}
    \Line(241,-216)(241,-219)
    \Line(225,-216)(241,-216)
    \Line(225,-219)(241,-219)
    \Line(225,-209)(225,-225)
    \Line(283,-216)(283,-219)
    \Line(283,-219)(299,-219)
    \Line(300,-209)(300,-225)
    \Line(283,-216)(299,-216)
    \DashCArc(262,-217)(21.1,-5.44,185.44){7}
    \COval(263,-196)(4,4)(0){Black}{White}
    \DashLine(241,-219)(283,-219){2}
  \end{picture}\\\nonumber
\end{eqnarray}
In the diagram above (and all those that follow) we use the graphic notation and the Feynman rules described in the Appendix, 
and denote with white (respectively black) blobs connected (respectively one particle irreducible) Green's functions.

Next, we need the BQI relating the trilinear vertex with all quantum fields 
to the vertex where a (quantum) scalar field has been replaced by a background one. 
To this end, we consider the functional differentiation
\begin{equation}
\left.\frac{\delta^3{\cal S'}(\Gamma')}{\delta A_\mu(k_1)\delta\Omega^{\phi}(q)\delta\phi^\dagger(p_1)}\right|_{\Phi=0}=0 \qquad q+p_1+k_1=0.
\end{equation}
Then, taking into account that, for the model at hand, all functions involving 
the combination $\Omega^\phi A^*_\mu$ are {\it a fortiori} 1PR, 
and therefore drop out from the identity, we find the identity
\begin{equation}
\Gamma_{A_\mu\phi^\dagger\widehat\phi}(p_1,q)=\left[1+\Gamma_{\Omega^\phi\phi^{*\dagger}}(q)\right]\Gamma_{A_\mu\phi^\dagger\phi}(p_1,q)+
\Gamma_{A_\mu\Omega^\phi\phi^{*\dagger}}(p_1,q)\Gamma_{\phi\phi^\dagger}(p_1).
\end{equation}
Introducing the notation $\widehat\Gamma_\mu\equiv\Gamma_{A_\mu\phi^\dagger\widehat\phi}$ and $G_\mu\equiv\Gamma_{A_\mu\Omega^\phi\phi^{*\dagger}}$, we can cast the above BQI in a short-hand form
\begin{equation}
\widehat\Gamma_\mu(p_1,q)=\left[1+G(q)\right]\Gamma_\mu(p_1,q)+
G_\mu(p_1,q)S^{-1}(p_1).
\label{BQIvert1}
\end{equation}
Notice that the function $G_\mu(q,p_1)$ has the simple expression
\begin{eqnarray}
iG_\mu(q,p_1)=\mbox{\hspace{3.2cm}}=g^2\int\![d\ell]\frac1{\ell^2}S(\ell+q)S(\ell-p_1)\Gamma_\mu(p_1-\ell,\ell+q).
  \begin{picture}(40,43) (557,-185)
    \SetWidth{0.5}
    \SetColor{Black}
    \Line(241,-191)(241,-194)
    \Line(225,-191)(241,-191)
    \Line(225,-194)(241,-194)
    \Line(225,-184)(225,-200)
    \Line(283,-191)(283,-194)
    \Line(283,-194)(299,-194)
    \Line(300,-184)(300,-200)
    \Line(283,-191)(299,-191)
    \DashCArc(262,-192)(21.1,-5.44,185.44){7}
    \COval(248,-178)(4,4)(0){Black}{White}
    \COval(277,-178)(4,4)(0){Black}{White}
    \GOval(263,-171)(4,4)(0){0.0}
    \DashLine(241,-194)(283,-194){2}
    \Photon(261,-142)(261,-171){1.5}{3.5}
    \Text(292,-207)[]{\small{\Black{$\Omega^\phi(q)$}}}
    \Text(235,-207)[]{\small{\Black{$\phi^{*\dagger}(p_1)$}}}
    \Text(283,-152)[]{\small{\Black{$A_\mu (p_2)$}}}
  \end{picture}\\\nonumber
\label{aux1}  
\end{eqnarray}

In order to obtain the BQI for the quadrilinear vertex 
one needs to consider the functional differentiation
\begin{equation} 
\left.\frac{\delta^4{\cal S'}(\Gamma')}{\delta A_\mu(k_1)\delta A_\nu(k_2)\delta\Omega^{\phi}(q)\delta\phi^\dagger(p_1)}\right|_{\Phi=0}=0 \qquad q+p_1+k_1+k_2=0,
\end{equation}
which provides the BQI
\begin{eqnarray}
\Gamma_{A_\mu A_\nu\phi^\dagger\widehat\phi}(k_2,p_1,q)&=&\left[1+\Gamma_{\Omega^\phi\phi^{*\dagger}}(q)\right]\Gamma_{A_\mu A_\nu\phi^\dagger\phi}(k_2,p_1,q)
+\Gamma_{A_\mu A_\nu\Omega^\phi\phi^{*\dagger}}(k_2,q,p_1)\Gamma_{\phi^\dagger\phi}(p_1)\nonumber \\
&+&\Gamma_{A_\mu\Omega^\phi\phi^{*\dagger}}(q,k_2+p_1)\Gamma_{A_\nu\phi^\dagger\phi}(p_1,-p_1-k_2)\nonumber\\
&+&\Gamma_{A_\nu\Omega^\phi\phi^{*\dagger}}(q,k_1+p_1)\Gamma_{A_\mu\phi^\dagger\phi}(p_1,-p_1-k_1).
\end{eqnarray}
Introducing the notation $\widehat\Gamma_{\mu\nu}\equiv\Gamma_{A_\mu A_\nu\phi^\dagger\widehat\phi}$ and $G_{\mu\nu}\equiv\Gamma_{A_\mu A_\nu\Omega^\phi\phi^{*\dagger}}$ we can rewrite the above BQI in its final form, namely
\begin{eqnarray}
\widehat\Gamma_{\mu\nu}(k_2,p_1,q)&=&\left[1+G(q)\right]\Gamma_{\mu\nu}(k_2,p_1,q)
+G_{\mu\nu}(k_2,q,p_1)S^{-1}(p_1)\nonumber \\
&+&G_\mu(q,k_2+p_1)\Gamma_\nu(p_1,-p_1-k_2)
+G_\nu(q,k_1+p_1)\Gamma_\mu(p_1,-p_1-k_1).
\label{BQIvert2}
\end{eqnarray}
The equation for the auxiliary function $G_{\mu\nu}$ is given by
\begin{eqnarray}
iG_{\mu\nu}(k_2,q,p_1)=\mbox{\hspace{3.2cm}}=g^2\int\![d\ell]\frac1{\ell^2}S(\ell+q)S(\ell-p_1)
{\cal C}_{\mu\nu}(k_2,p_1-\ell,\ell+q),
  \begin{picture}(10,43) (573,-185)    
  \SetWidth{0.5}
    \SetColor{Black}
    \Photon(266,-171)(242,-142){-1.5}{3.5}
    \Photon(261,-171)(285,-142){1.5}{3.5}
    \Text(295,-207)[]{\small{\Black{$\Omega^\phi(q)$}}}
    \Text(235,-207)[]{\small{\Black{$\phi^{*\dagger}(p_1)$}}}
    \Line(241,-191)(241,-194)
    \Line(225,-191)(241,-191)
    \Line(225,-194)(241,-194)
    \Line(225,-184)(225,-200)
    \Line(283,-191)(283,-194)
    \Line(283,-194)(299,-194)
    \Line(300,-184)(300,-200)
    \Line(283,-191)(299,-191)
    \DashCArc(262,-192)(21.1,-5.44,185.44){7}
    \COval(248,-178)(4,4)(0){Black}{White}
    \COval(277,-178)(4,4)(0){Black}{White}
    \COval(263,-171)(4,4)(0){Black}{White}
    \DashLine(241,-194)(283,-194){2}
    \Text(294,-159)[]{\small{\Black{$A_\nu(k_2)$}}}
    \Text(235,-159)[]{\small{\Black{$A_\mu(k_1)$}}}
  \end{picture}\nonumber\\
\label{aux2}
\end{eqnarray}
where ${\cal C}_{\mu\nu}$ is the 
 four-particle connected Green's function with 
two photons and two scalar entering, whose properties and 
WI will be discussed in detail later.

Finally, we report for completeness  
the WI satisfied by the auxiliary functions $G_{\mu}$ and
$G_{\mu\nu}$. Contracting directly 
their defining equations, Eq.~(\ref{aux1}) and Eq.~(\ref{aux2}), using Eq.~(\ref{WI3}), and
the WI for the kernel ${\cal C}_{\mu\nu}$ derived in Section \ref{ck}, Eq.~(\ref{kWI1}),
we obtain
\bea
p_2^{\mu}\, G_{\mu}(q,p_1) &=&  g\,[G(q) - G(p_1)], \nonumber\\ 
k_1^{\mu}\, G_{\mu\nu}(k_2,q,p_1) &=& g\, [G_{\nu}(q+k_1,p_1) - G_{\nu}(q,p_1+k_1)].
\label{auxWI}
\eea

\section{\label{ptpert} PT in scalar QED: General considerations}

In this section we present the general  methodology for constructing PT self-energies 
in the case of scalar QED. The object of interest will be the 
scalar self-energy: 
due to the momenta appearing in the elementary vertices, the PT algorithm allows
the conversion of the standard scalar self-energy into the 
BFM scalar self-energy, {\it e.g.}, with a background scalar entering and exiting.
Note that the photon self-energy remains intact, because 
no WI can be triggered within the corresponding graphs defining it.
After outlining the general philosophy and
setting up some useful notation, we will proceed to 
review the construction of PT scalar self-energy at one and two loops.

The general  idea of the  PT is to  identify, following a  very strict
procedure,  propagator-like  contributions  contained in  vertex-  and
box-diagrams,  and  reassign  them  to  the  conventional  self-energy
graphs~\cite{Cornwall:1982zr}, thus generating new, effective Green's functions, with special
properties.   This  construction is  carried  out  inside an  $S$-matrix
element,  or  some other  gauge-invariant  observable; the  underlying
symmetries,   most  notably   the  BRST   symmetry,   enforce  crucial
cancellations, making  the aforementioned construction  possible.  The
``$S$-matrix'' PT  described above has  an equivalent version,  known as
``intrinsic'' PT~\cite{Cornwall:1989gv}.
According to  it, one identifies the parts
of the self-energy that will cancel against the pinching terms coming
from  vertices  and  boxes,   and  discards  them  directly  from  the
self-energy:  what  remains  is   the  answer.  The  intrinsic  PT  is
operationally more economical, 
and minimizes the need of embedding the
procedure into  a physical observable;  in what follows we  will adopt
this latter approach.

The rearangments of graphs mentioned above are realized when  
judiciously selected longitudinal momenta, 
circulating inside the Feynman graphs, trigger elementary WI's.
These momenta stem either from the 
longitudinal (gauge-dependent) parts of the propagators, or 
from parts of the momenta carried by the ``external'' elementary vertices~\cite{Papavassiliou:1999az}
{\it i.e.}, vertices where the physical momentum enters or leaves the corresponding diagram. 
The construction is simplified enormously 
in the context of the renormalizable gauges, if one chooses 
directly the Feynman gauge~\cite{Cornwall:1989gv}. This choice eliminates all pinching momenta,
other than those stemming from the external vertices. 
Such a choice constitutes no loss of generality, as has been explained in~\cite{Binosi:2004qe},
by establishing a close correspondence between the PT and
the powerful  Nielsen identities~\cite{Nielsen:1975fs} that control 
the gauge-dependence of the conventional Green's functions.    

\begin{figure}[t]
\includegraphics[width=14cm]{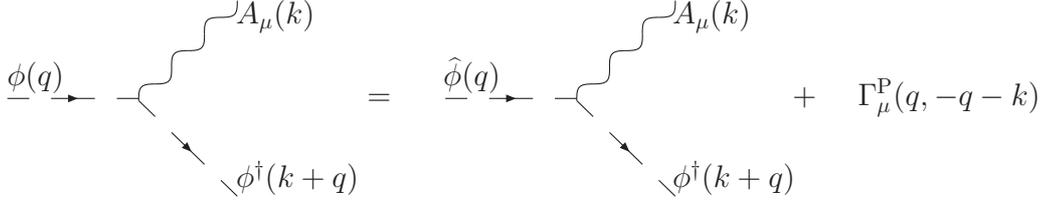}
\caption{\it The PT decomposition of the elementary vertex}
\label{PTsplit}
\end{figure}

Let us now turn to the case of scalar QED, and outline the construction of the 
PT scalar self-energy. 
The  bare  photon   propagator, $\Delta_{\mu\nu}^{(0)}(k)$, assumes the form
\be
\Delta_{\mu\nu}^{(0)}(k) = 
-\frac{i}{k^2}
\left[\ g_{\mu\nu} - (1-\xi) \frac{k_\mu k_\nu}{ k^2}\right] \,,
\label{PhotProp}
\ee 
and, following the previous discussion,
 we will choose directly the Feynman gauge, $\xi=1$. 
Therefore, the only pinching momenta will originate from 
the bare (tree-level) scalar-scalar-photon vertex $\Gamma^{(0)}_\mu$.
According to the PT, this latter vertex
is to be split into two parts:
({\it i}) a part, to be denoted by 
 $\Gamma^{\mathrm P}$, which contains  
longitudinal momenta, {\it i.e.}, momenta that 
can be contracted with the vertex on the other side of the diagram,
thus triggering an elementary WI, and  
({\it ii}) the remainder, to be denoted by $\Gamma^{\mathrm F}$,
which coincides with the corresponding tree-level vertex in the BFM; in particular,
the background field is to be identified with the field carrying the external momentum.
In the case of the scalar vertex $\Gamma^{(0)}_\mu(q,-q-k)$,
the only longitudinal momentum is $k_\mu$, irrigating the photon line; therefore,    
the PT decomposition of the vertex described above amounts to (see also Fig.\ref{PTsplit})
\be
\Gamma^{(0)}_\mu(q,-q-k)=
\Gamma^{\mathrm F}_\mu(q,-q-k)+\Gamma^{\mathrm P}_\mu(q,-q-k), 
\label{PTsplit1}
\ee
 with 
\begin{eqnarray}
i\Gamma^{(0)}_\mu(q,-q-k) &=& -ig(2q+k)_\mu \,,\nonumber\\
i\Gamma^{\mathrm F}_\mu(q,-q-k) &=& -2ig q_{\mu} \,,\nonumber\\
i\Gamma^{\mathrm P}_\mu(q,-q-k) &=& -igk_\mu \,.
\label{PTsplit2}
\end{eqnarray}

For the case of the scalar self-energy, 
the above splitting is to be carried out to the two external vertices, 
where the physical momentum $q$ is entering and exiting: Specifically, we write
\be
\Gamma^{(0)}_\mu [...]\Gamma^{(0)}_\nu = \Gamma^{\mathrm F}_\mu [...]\Gamma^{\mathrm F}_\nu
+\Gamma^{\mathrm P}_\mu [...]\Gamma^{(0)}_\nu
+\Gamma^{(0)}_\mu [...]\Gamma^{\mathrm P}_\nu-\Gamma^{\mathrm P}_\mu [...]\Gamma^{\mathrm P}_\nu.
\label{PTsplit3}
\ee
where $[...]$ denotes the rest of the diagram appearing 
between the two vertices.
In what follows we will use the short-hand notation
$[dk] = \mu^{\epsilon}d^d k/(2\pi)^d$, 
with $d=4-\epsilon$, the dimension of space-time, and $\mu$ the 't Hooft mass; also, we will use roman letters to label Feynman diagrams  computed in the $R_\xi$ gauge, and roman letters with hats when the same diagram is computed in the BFM gauge.
For the perturbative analysis of this section, 
we will employ the scalar self-energy, $\Sigma$, related to the 
inverse scalar propagator by  
\be
S^{-1}(q) = q^2-m^2 + i \Sigma(q) , 
\ee
and the same relation applies for ${\widehat S}^{-1}$ and 
${\widehat \Sigma}$.
In addition, we will use  $S_{0}^{-1}(q)= q^2-m^2$. 

\subsection{One-loop case}

\begin{figure}[!t]
\includegraphics[width=14cm]{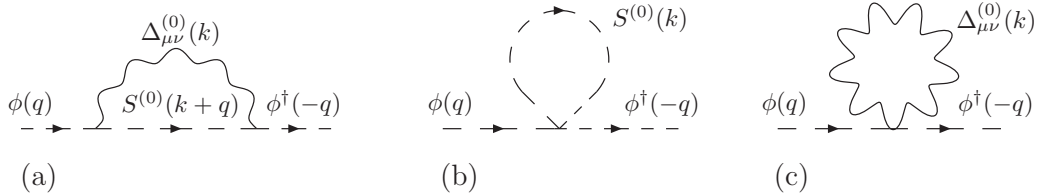}
\caption{\it The diagrams which, when evaluated using the corresponding Feynman rules
(see Appendix, Fig.\ref{fr_sqed}), contribute 
at one loop to the conventional and BFM scalar self-energies, $\Sigma^{(1)}(q)$ 
and ${\widehat\Sigma}^{(1)}(q)$, respectively.}
\label{1loop}
\end{figure}

The one-loop case is particularly  simple. In fact, recall that we are
working in the Feynman gauge; then, since graph (b) of Fig.\ref{1loop}
can not possibly provide any pinching momenta, while graph (c) is zero
in perturbation theory [due to Eq.(\ref{dim_reg}) below], one needs to
concentrate only on  diagram (a). Then, by applying  to the latter the
decomposition described in Eq.(\ref{PTsplit3}),  one should be able to
generate graph $(\widehat{\mathrm{a}})$, together  with the rest of the
terms   appearing   in   the   one-loop   version  of   the   BQI   of
Eq.(\ref{BQIfinal}), namely
\be    
\Sigma^{(1)}(q)   = {\widehat\Sigma}^{(1)}(q)  -  2  G^{(1)}(q)S^{-1}_{0}(q).  
\ee  
Notice
that,  in  the  one  loop   case,  the  symbol  $[...]$  appearing  in
Eq.(\ref{PTsplit3})  is  given  by  the  expression  $-  i  g_{\mu\nu}
S_{0}(q+k)/k^2$.  In  what follows  we  will  denote symbolically  the
application  of  Eq.  (\ref{PTsplit3})  on (a)  as  \be  (\mathrm{a})=
(\mathrm{a})^{\mathrm{F}\mathrm{F}}   +  (\mathrm{a})^{\mathrm{P}0}  +
(\mathrm{a})^{0\mathrm{P}} - (\mathrm{a})^{\mathrm{P}\mathrm{P}}.
\label{gendef}
\ee
The notation introduced above will be used extensively in the rest of the paper.

The term $(\mathrm{a})^{\mathrm{F}\mathrm{F}}$ on the RHS of (\ref{gendef})
represents already graph $(\widehat{\mathrm{a}})$ of Fig.\ref{1loop}, since 
$\Gamma^{(0)}_\mu(q,-q-k)$ coincides (by construction) with the tree-level 
$\widehat{\phi}\phi^\dagger A$ vertex. Then, using the tree-level version of the all-order WI given in Eq.(\ref{WI3}), 
it is elementary to demonstrate that the second and third term on the RHS of (\ref{gendef})
give each rise to a term $S_{0}^{-1}(q)G^{(1)}(q)$, with $G^{(1)}$ the one-loop 
version of Eq.(\ref{aux-function}). The last term, $(\mathrm{a})^{\mathrm{P}\mathrm{P}}$, gives rise to a seagull-like graph in which the four scalar vertex is proportional to the gauge coupling $g^2$; it is this latter term combined with diagram (b) that will give rise to the
characteristic BFM vertex $\phi^\dagger\phi\widehat{\phi}^{\dagger}\widehat{\phi}\propto(\lambda-g^2)$ (see the Appendix for its exact Feynman rule) and therefore to the diagram ($\widehat{\mathrm{b}}$).

We end by observing that in carrying out the construction above we have used the dimensional regularization result 
$\int\![dk]/k^2=0$, a special case of the more general formula
\be
\int \!\! \frac{[dk]}{k^2} \ln^{N} (k^2)  =0 \,, \quad N =0, 1,2,\ldots
\label{dim_reg}
\ee
which guarantees the masslessness of the photon to all orders in perturbation theory [and that graph (c) is zero as well].

\subsection{Two-loop case}

The two-loop case is of course more involved; in fact, it has sufficient level of complexity  
to capture all central issues one needs to address for the 
all-order perturbative construction, as well as the generalization 
at the level of the SDEs, to be presented in the next sections.

As in the one-loop case, the idea is again to start out with the 
graphs defining the conventional two-loop scalar self-energy $\Sigma^{(2)}$, and to generate, 
via the application of the PT rules, the diagrams of 
the corresponding two-loop BFM self-energy ${\widehat\Sigma}^{(2)}$ together with all additional terms
enforcing the BQI of (\ref{BQIfinal}) at two loops.
In particular, the two-loop version of Eq.(\ref{BQIfinal}) is given by
\be
\Sigma^{(2)}(q) =  {\widehat\Sigma}^{(2)}(q) - 2 G^{(1)}(q) \Sigma^{(1)}(q) -
2  G^{(2)}(q) S^{-1}_{0}(q) - [G^{(1)}(q)]^2 S^{-1}_{0}(q).
\label{BQI2l}
\ee
Before entering into the details, we report the form of the all-order 
photon propagator, $\Delta_{\mu\nu}(k)$, in the Feynman gauge. We have 
\be
\Delta_{\mu\nu}(k) = -i\left[\Delta(k^2)P_{\mu\nu}(k)+\frac{k_\mu
k_\nu}{k^4}\right], \qquad \Delta(k^2)=\frac 1{k^2+i\Pi(k^2)},
\label{Dgen}
\ee
where $P_{\mu\nu}(k)=g_{\mu\nu} - k_\mu k_\nu/k^2$ denotes the 
dimensionless projection operator, and
\mbox{$\Pi_{\mu\nu}(k)=\Pi(k^2)P_{\mu\nu}(k)$} is the transverse vacuum polarization.

\begin{figure}[!t]
\includegraphics[width=15cm]{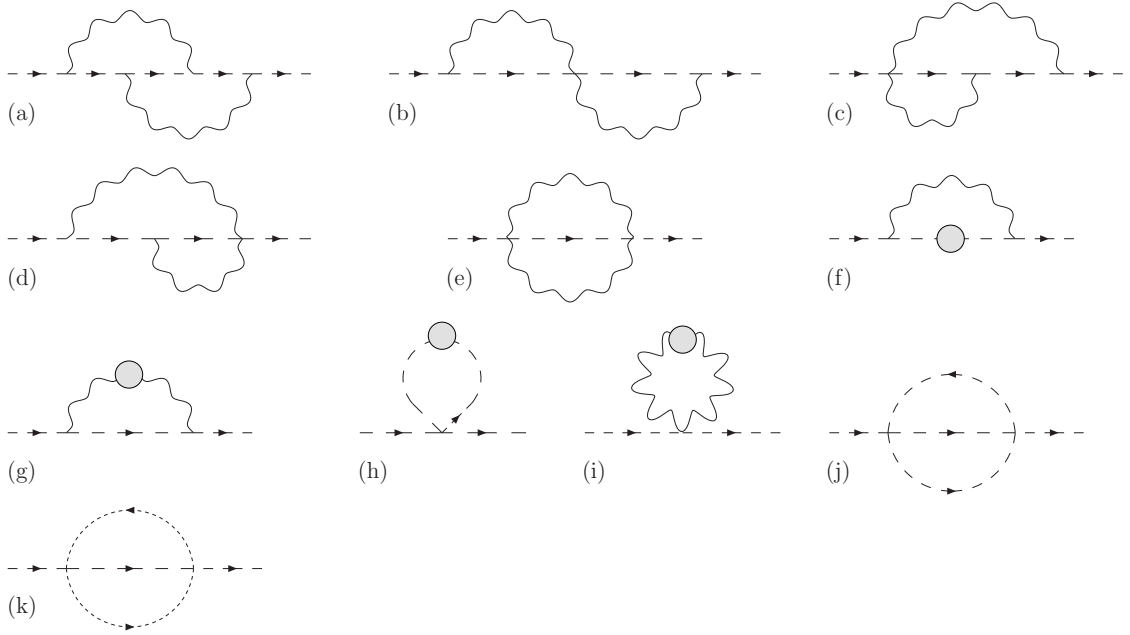}
\caption{\it The two-loop Feynman diagrams contributing to ${\Sigma}^{(2)}$ and 
${\widehat\Sigma}^{(2)}$: the topologies 
$(\mathrm{a})$--$(\mathrm{j})$, evaluated using the corresponding set of Feynman rules, contribute to both,
whereas $(\mathrm{k})$ only to the latter. Gray blobs represent the one-loop scalar and photon self-energy}
\label{2loop}
\end{figure}

The two-loop PT construction proceeds then as follows.

\begin{itemize}

\item[({\it i})]  $\Sigma^{(2)}$ is given by the sum of 
the one-particle irreducible (1PI) diagrams 
(a)--(j), shown in Fig.\ref{2loop};
on the other hand, ${\widehat\Sigma}^{(2)}$ is given by the sum  (a)--(k) 
(in each case one would be of course using the corresponding set of Feynman rules). 
Notice in particular that diagram (k) 
is due to the characteristic BFM vertex 
$\widehat{\phi}\phi^{\dagger}\bar{c}c$, shown in the Appendix.

\item[({\it ii})] Diagrams ({a})--({j}) of $\Sigma^{(2)}$ may be classified 
into three categories, according to the number of  external
$\phi^{\dagger}\phi A$ vertices they contain:
type A diagrams with two such vertices --  
graphs ({a}), ({b}), ({f}), and ({g}); type B diagrams with one such vertex --
graphs ({c}) and (d); type C diagrams with no such vertex -- 
graphs ({e}), ({h}) ({i}) and ({j}).
Then, to the type A diagrams one applies the rearrangement given in Eq.(\ref{PTsplit3}), whereas 
for the type B one simply carries out the PT splitting 
of Eq.(\ref{PTsplit1}) to their single 
external vertex. Finally, type C diagrams remain unchanged, as they 
do not contain any pinching momentum.

\item[({\it iii})] In type A graphs, the terms containing  
$\Gamma^{\mathrm F} [...]\Gamma^{\mathrm F}$ give rise to the corresponding
BFM diagrams, to be denoted by  $(\widehat{\mathrm{a}})$, $(\widehat{\mathrm{b}})$, 
$(\widehat{\mathrm{f}})$, and $(\widehat{\mathrm{g}})$.
Similarly, the terms containing $\Gamma^{\mathrm F}$ in type B graphs
generate the corresponding BFM diagrams $(\widehat{\mathrm{c}})$ and $(\widehat{\mathrm{d}})$.
Thus, 
\bea
(\mathrm{x})^{\mathrm{F}\mathrm{F}} &=&  (\widehat{\mathrm{x}}) \qquad \mathrm{x}= 
\mathrm{a}, \mathrm{b}, \mathrm{f}, \mathrm{g} \nonumber\\
(\mathrm{y})^{\mathrm{F}} &=& (\widehat{\mathrm{y}}) \qquad \mathrm{y}=\mathrm{c}, \mathrm{d}
\eea
Notice that, due to the transversality of the photon self-energy, $k^{\mu}\Pi_{\mu\nu}^{(1)}(k)=0$, 
diagram ({g}) gives no further contributions, i.e. it has been converted into the corresponding diagram 
$(\widehat{\mathrm{g}})$ for free. As for diagram ({f}), using the 
{\it tree-level} version of Eq.(\ref{WI3}), we have:
\bea
(\mathrm{f})^{\mathrm{P}0}+(\mathrm{f})^{0\mathrm{P}} &=& 
2\, g^2 \int \!\frac{[dk]}{k^2}S_0(k+q)\, \Sigma^{(1)}(k+q) - 2 G^{(2)}(q) S^{-1}_{0}(q) 
\nonumber\\
(\mathrm{f})^{\mathrm{P}\mathrm{P}} &=& - g^2\int \![dk]\,\Sigma^{(1)}(k).
\label{ff}
\eea

\item[({\it iv})] We continue with the evaluation of the pinching parts 
of the remaining graphs considered in ({\it iii}).
Combining the two type A graphs ({a}) and ({b}) with the two type B ones, 
we may organize the various contributions such that the two 
$\Gamma^{\mathrm P}$ are each acting on the {\it full} one-loop 
$\phi^{\dagger}\phi A$ vertex $\Gamma^{(1)}_{\mu}$, thus triggering the one-loop
version of the WI of Eq.(\ref{WI3}). 
Thus, the $\Gamma^{\mathrm P}$ on the left hand side (LHS)
of diagrams ({a}), ({b}), and ({d}), will act on $\Gamma^{(1)}_{\mu}$,
and exactly the same will happen with the $\Gamma^{\mathrm P}$ on the right hand side (RHS)
of diagrams ({a}), ({b}), and ({c}).
Specifically,
\bea
2 \left[(\mathrm{a})^{\mathrm{P}0} + (\mathrm{b})^{\mathrm{P}0} + (\mathrm{d})^{\mathrm{P}}\right] &=& 
2 g \int \!\frac{[dk]}{k^2}S_0(k+q)\, k^{\mu}\Gamma_{\mu}(-q,k+q)
\nonumber\\
&=& 
-2g^2 \int \!\frac{[dk]}{k^2}S_0(k+q)\Sigma^{(1)}(k+q)-2 G^{(1)}(q) \Sigma^{(1)}(q), \nonumber \\ 
\label{abd}
\eea
where the multiplicative factor of 2 accounts for the equal contribution from the symmetric combination
$(\mathrm{a})^{0\mathrm{P}} + (\mathrm{b})^{0\mathrm{P}} + (\mathrm{c})^{\mathrm{P}}$.
Finally, it is straightforward to demonstrate that 
\be
(\mathrm{a})^{\mathrm{P}\mathrm{P}}+(\mathrm{b})^{\mathrm{P}\mathrm{P}} = - (\mathrm{k}) + \left[G^{(1)}(q)\right]^2S_0^{-1}(q). 
\label{pp}
\ee
Evidently, diagram ({k}), originating from the special BFM ghost sector, 
has been generated dynamically from the rearrangement of diagrams 
evaluated with Feynman rules that do not involve ghost interactions. 
[To get the signs to work out, remember the minus sign in front of the 
$\Gamma^{\mathrm P} [...]\Gamma^{\mathrm P}$ term, and the extra
minus sign in ({k}) due to the ghost loop.]

\item[({\it v})] Finally, taking into account the cancellation of the first terms on the RHS of 
Eqs.(\ref{ff}) and (\ref{abd}), we conclude that all diagrams contributing to 
${\widehat\Sigma}^{(2)}$ have been generated by pinching internally
$\Sigma^{(2)}$, together with all terms on the RHS of (\ref{BQI2l}); this concludes the 
two-loop construction of the PT scalar self-energy.

\end{itemize}

As has been explained in detail in the literature (for the more complicated non-Abelian case)~\cite{Papavassiliou:1999az}, 
all terms in Eq.(\ref{BQI2l}) containing the auxiliary function $G$ will eventually cancel
exactly in an  $S$-matrix element (or other physical observable) against similar 
contributions coming from the conversion of the two-loop conventional vertex 
$\Gamma^{(2)}_{\mu}$ to the PT-BFM vertex ${\widehat\Gamma}^{(2)}_{\mu}$, 
[{\it viz.} Eq.(\ref{BQIvert1})], together 
with analogous terms originating from the conversion of  
the 1PR strings 
(i.e. products of conventional one-loop 
vertices and self-energies) into PT strings, {\it e.g.}, 
1PR strings containing instead products of one-loop PT vertices and self-energies. 

\section{\label{sdq} Pinching Schwinger-Dyson equations}

We now enter into the main issue of this article, namely how to carry out 
the PT construction at the level of SDE. In this section we will 
present a general qualitative discussion of the main questions involved, the strategy 
that will be employed, and the field-theoretic ingredients necessary for its 
implementation. The actual detailed construction will be presented in Section \ref{sde}.

\begin{figure}[!t]
\includegraphics[width=15cm]{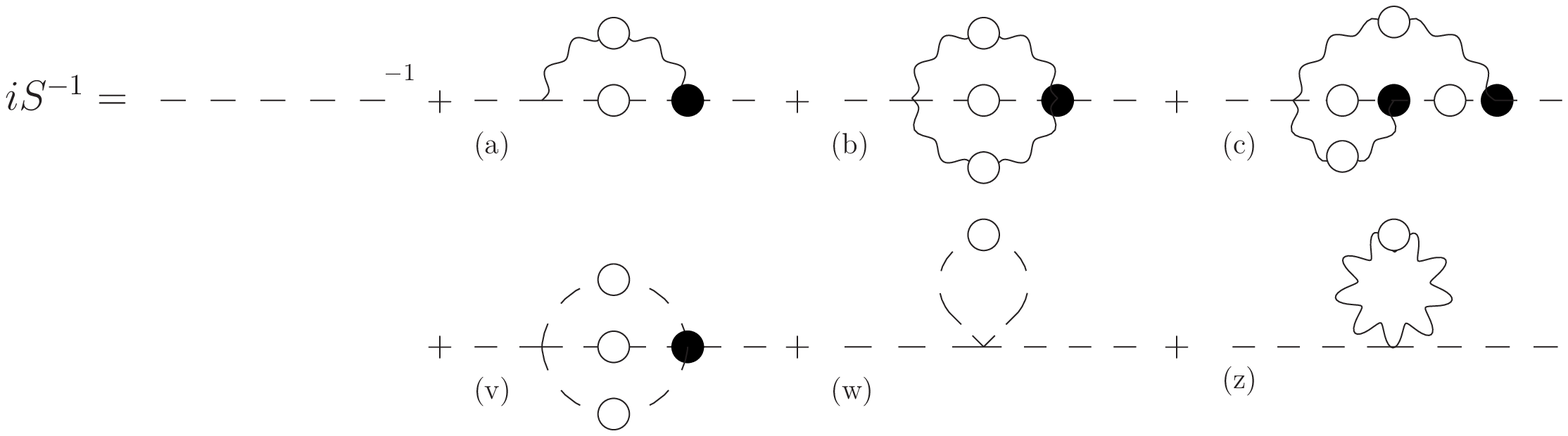}
\caption{\it Schwinger-Dyson equations for the scalar self-energy 
$iS^{-1}(q)$. The charge flow is not shown. Here black blobs represent 1PI Green's functions ($\Gamma_\mu$, $\Gamma_{\mu\nu}$ and $\Gamma$), and white blobs connected Green's functions ($\Delta_{\mu\nu}$ and $S$).}
\label{SDE-prop}
\end{figure}

The SDE may be derived following a diagrammatic analysis in the spirit
of~\cite{Bjorken:1979dk}, or  formally from the  generating functional
of the theory, as  shown, for example, in~\cite{Itzykson:1980rh}.  For
the case of scalar QED, the SDE for the scalar propagator $S$ is shown
in Fig.\ref{SDE-prop}; it essentially  amounts to dressing with vertex
and self-energy  corrections all skeleton graphs  contributing to $S$.
In  scalar  QED  the  skeleton  graphs  of  the  scalar  (and  photon)
propagator are  exhausted at two  loops; this is equivalent  to saying
that, with the elementary vertices  at hand, any higher order graph is
bound to be a radiative correction (propagator or vertex ``dressing'')
to the one-  and two-loop graphs.  That this is so  may be verified by
direct  diagrammatic  analysis,  analogous  to the  one  presented  in
\cite{Bjorken:1979dk}  for the  case of  spinor QED.  In  general, the
number of skeleton  graphs depends on the type  of elementary vertices
characterizing  the  theory.  Thus,  in  spinor  QED  the fermion  and
electron self-energies  have a single skeleton graph,  the analogue of
graph (4a); the same is true for a $g\phi^3$ theory. In both cases the
reason is  that there exists only one  fundamental interaction vertex.
Instead, in the case of a $g\phi^3 + \lambda \phi^4$ theory, of scalar
QED, and of QCD, due to the presence of interaction vertices involving
four fields, two-loop skeleton graphs exist; for example, no radiative
correction to graph  (2a) could possibly give rise  to the graphs (3c)
and  (3e). Clearly, field theories with  elementary vertices 
involving more than four  fields have 
self-energy skeleton  graphs beyond  two loops.

\begin{figure}[!t]
\includegraphics[width=15cm]{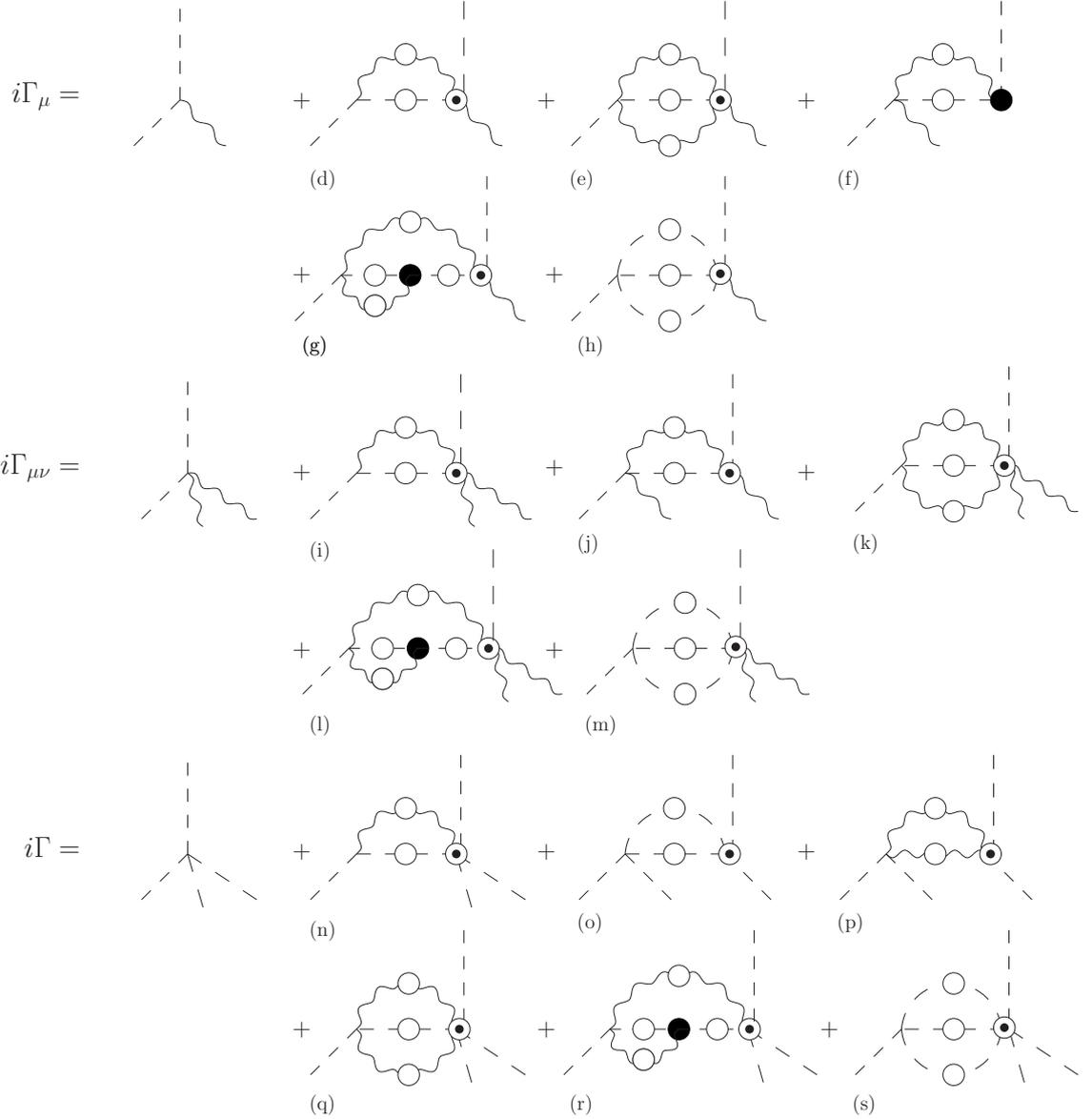}
\caption{\it Schwinger-Dyson equations for the three fundamental vertices, $i\Gamma_{\mu}(p_1,p_2)$, $i\Gamma_{\mu\nu}(k,p_1,p_2)$ and 
$i\Gamma(p_1,p_2,p_3)$. As before, black blobs represent 1PI Green's functions ($\Gamma_\mu$), white blobs connected 
Green's functions ($\Delta_{\mu\nu}$ and $S$), and white blobs with a black center denote the 
various kernels.}
\label{SDEs-vert}
\end{figure}

There are  three fundamental
(fully  dressed, all-order)  vertices  appearing in  the  SDE of  $S$,
corresponding   to  the   couplings  $\partial_{\mu}\phi^{\dagger}\phi
A^{\mu}$, $\phi^{\dagger}\phi A^{\mu}A_{\mu}$, and $\phi^{\dagger}\phi
\phi^{\dagger}\phi$,     to    be    denoted     by    $\Gamma_{\mu}$,
$\Gamma_{\mu\nu}$, and $\Gamma$, respectively.  The corresponding SDEs
are  shown in  Fig.\ref{SDEs-vert}.   Their general  structure may  be
described as  follows: A  vertex-leg is singled  out (in our  case the
scalar  leg  carrying  momentum  $q$),  and  all  possible  tree-level
vertices involving this field (leg) are written down. Then, the fields
exiting  from  these tree-level  vertices  are  either  ({\it i})  all
connected   with  the   remaining   vertex-legs  through   appropriate
multi-particle  kernels,  or  ({\it  ii})  one  of  them  is  directly
identified with one of the  vertex-legs, whereas the rest is connected
to  the remaining vertex-legs  through an  appropriate kernel  or full
vertex. The  various kernels involved  (to be  denoted in
what follows  as ${\cal K}_{\mu\nu}$ and  ${\cal K}_{\mu\nu\rho}$) are
connected, and, in  addition, 1PI with respect to  cuts involving only a
{\it physical}  momentum; this is tantamount to saying 
that these kernels  do not contain graphs that could become disconnected 
by cutting a single line
that is irrigated exclusively by one of the external, (``physical'' as
opposed to ``virtual'') momenta entering into the vertex.  
It is important to emphasize that there is a finite number of distinct 
$n$-particle kernels appearing in the SDE for the 
vertices of Fig.\ref{SDEs-vert}. Specifically, 
the SDE of a vertex with 
$m$-fields ($m=3,4$, in our case) will involve all kernels with  
$n \leq m+2$. 
To see with an example why this must be so, 
consider the  SDE of $\Gamma_{\mu}$ in Fig.5, and let us add 
one additional leg to the 5-particle kernel appearing either in (e) or in (h).
Since the number of external legs is fixed, this extra leg must be
attached to the rest of the diagram through an internal elementary vertex.
The resulting graph, however, will be nothing but a 
radiative correction to one of the graphs containing
the kernels with $n \leq 5$; therefore, its inclusion would 
constitute overcounting. 
Notice finally
that we do not consider the  SDE for the photon propagator, because in
scalar QED  it remains completely  inert in the PT  rearrangement (the
most  direct  way  to see  this  is  by  noticing that  all  couplings
involving  a  conventional photon  coincide  with  those containing  a
background photon).  

The key observations that allow for the extension
of the PT algorithm at the level of the SDE of the theory are then the
following.
 
\begin{itemize}

\item[({\it i})] In order to carry out the PT construction for the SDE at hand, it is important to 
first identify the origin of the pinching momenta ({\it i.e.}, type A and B diagrams), and then 
the structures (vertices, kernels, etc) these momenta
will be acting upon. Let us focus for concreteness 
on the SDE for the scalar propagator $S$.
To determine the pinching momenta, we apply the same criterion as in the perturbative case,
namely we carry out the PT decomposition to the ``external'' vertices.  
Looking at the diagrams of Fig.\ref{SDE-prop}, 
it is clear that the bare vertex on the very left of diagram 
({a}) should 
be decomposed according to Eq.(\ref{PTsplit1}); what is less clear perhaps is how to 
implement the subsequent splitting described in Eq.(\ref{PTsplit3}), 
or in other words, identify the second external vertex to be decomposed. 
The perturbative examples studied in the previous section suggest  
that the second external vertex resides inside the black blob denoting the full trilinear vertex $\Gamma_\mu$ in ({a});
thus, in order to implement  Eq.(\ref{PTsplit3}) one must ``unwrap'' $\Gamma_\mu$. 
This is accomplished by 
considering the SD for $\Gamma_\mu$ itself (see Fig.\ref{SDEs-vert}), 
which contains indeed tree-level vertices $\Gamma_\mu^{(0)}$,
on its RHS (the first two terms). Then, one must    
think of the RHS of this latter SDE 
as having been inserted in ({a}), instead of $\Gamma_\mu$,
and carry out Eq.(\ref{PTsplit1}) on the tree-level vertices $\Gamma_\mu^{(0)}$ 
now appearing explicitly on the right of ({a}), see, {\it e.g.}, Fig.\ref{diags-abc}. 

\item[({\it ii})] After having settled the question of how to identify type A  
diagrams and how to carry out the implementation of Eq.(\ref{PTsplit3})
in the presence of a full trilinear vertex, the next question is 
what the result of this operation will be, and in particular the 
contributions of the terms $\Gamma^{\mathrm P}_\mu [...]\Gamma^{(0)}_\nu$, 
$\Gamma^{(0)}_\mu [...]\Gamma^{\mathrm P}_\nu$, 
and $\Gamma^{\mathrm P}_\mu [...]\Gamma^{\mathrm P}_\nu$.
It is clear that the $\Gamma^{\mathrm P}$ originating from  the
tree-level vertex on the left of ({a}) will trigger {\it directly}
the WI of Eq.(\ref{WI3}), since it acts on a full $\Gamma_{\mu}$.
The result of the action of the $\Gamma^{\mathrm P}$ 
coming from the other side is, however, less transparent. Of course, 
our perturbative experience tells us that $\Gamma^{\mathrm P}_\mu [...]\Gamma^{(0)}_\nu$
and 
$\Gamma^{(0)}_\mu [...]\Gamma^{\mathrm P}_\nu$ should give identical contributions;
however, unlike the perturbative case where the symmetry of the situation
is manifest, now one has to {\it demonstrate} that this is indeed the case.
The way this is done is by noticing that, just as happened in the two-loop case where
$(\mathrm{a})^{0\mathrm{P}}$ was combined with 
$(\mathrm{b})^{0\mathrm{P}}$ and $(\mathrm{c})^{\mathrm{P}}$ to generate $\Gamma^{(1)}_{\mu}$,
now one has to consider the analogous contributions from graphs 
$(\mathrm{b})$, $(\mathrm{c})$, and $(\mathrm{w})$ of Fig.\ref{SDE-prop}. Specifically,
one must carry out the PT decomposition on the corresponding 
full vertices (black ``blobs''),
appearing on the very right of these graphs; this is accomplished again by 
unwrapping them, invoking their own SDE's, 
and carrying out  Eq.(\ref{PTsplit1}) on the tree-level trilinear vertices appearing 
on their RHS. The end-result of this will be that the $\Gamma^{\mathrm P}$ 
coming from the right will be acting on a set of diagrams that will be 
{\it precisely} the RHS of the SDE for  $\Gamma_{\mu}$; thus, the WI of Eq.(\ref{WI3})
will be triggered again, as expected.

\item[({\it iii})] Next we turn to the term $\Gamma^{\mathrm P}_\mu [...]\Gamma^{\mathrm P}_\nu$.
It should be clear from the two-loop construction, that 
what is contained in [...] of ({a}) and ({b}) in Fig.\ref{2loop}
is nothing but the tree-level 1PI kernel 
containing two scalars and two photons, {\it i.e.}, ${\cal K}^{(0)}_{\mu\nu}$.
So, what one is actually computing at two-loop in order to arrive at 
Eq.(\ref{pp}) is the tree-level WI for ${\cal K}_{\mu\nu}$.
This observation persists at the level of the propagator SDE: to determine 
$\Gamma^{\mathrm P}_\mu [...]\Gamma^{\mathrm P}_\nu$ one must 
find the result of fully contracting  ${\cal K}_{\mu\nu}$ by the 
corresponding momenta carried by the two photons entering. 

\item[({\it iv})] Turning to the SD equations for the vertices, let us first observe that the 
PT procedure can be implemented by simply carrying out the PT 
decomposition to the corresponding  vertices $\Gamma^{(0)}_{\mu}$
appearing on the corresponding RHS. Thus, for the case of  $\Gamma_{\mu}$ 
one must decompose the $\Gamma^{(0)}_{\mu}$ appearing in graph ({d}),
and determine the action of the longitudinal momentum on the 
kernel ${\cal K}_{\mu\nu}$. For the case of  $\Gamma_{\mu\nu}$
one must do the same in graph ({i}), and thus determine the action
of the longitudinal momentum coming from $\Gamma^{\mathrm P}_\mu$ on the 
five-particle (three photons and two scalars) kernel ${\cal K}_{\mu\nu\rho}$.

\item[({\it v})] Let us also emphasize that, for the purpose of pinching the propagator SDE 
{\it alone}, one does not need the WIs for the multi-particle kernels 
appearing in the various SDEs, other than ${\cal K}_{\mu\nu}$. 
Indeed, as has been outlined above, when the relevant contributions from the vertices 
are inserted into the SDE of $S$, and are appropriately combined with 
other graphs, the pinching momentum acts finally on a full $\Gamma_{\mu}$,
triggering its known WI.
The need for the WI satisfied by  ${\cal K}_{\mu\nu\rho}$, etc.  arises only
if one decides to pinch in addition the SDEs for  $\Gamma_{\mu}$, $\Gamma_{\mu\nu}$, and $\Gamma$.
In Section \ref{sde} 
we will pinch the SDE for $\Gamma_{\mu}$ and $\Gamma_{\mu\nu}$,
but will skip the case of the four-scalar vertex $\Gamma$;
the latter is straightforward but tedious, and 
presents limited conceptual or practical interest. Thus, 
the only WIs needed for our purposes are those  
for the kernels ${\cal K}_{\mu\nu}$ and ${\cal K}_{\mu\nu\rho}$.

\end{itemize}

Summarizing, we have seen that the PT construction can be carried out at the level of the SDE,
when appropriate adjustments to the perturbative methodology are implemented.
In particular, in the construction of the 
PT scalar self-energy $S(q)$, in addition 
to its own SDE, one must simultaneously consider the SDE for the full 
vertices involved, manipulating them appropriately. Furthermore,
it has become clear that one needs to derive closed expressions for the 
all-order WI satisfied by 1PI multi-particle kernels. This question will be addressed in 
detail in the next section.

\section{\label{ck} Ward identities for kernels}

In this section we will derive the WI needed for 
carrying out the PT construction for the SDEs of $S$, $\Gamma_{\mu}$, and 
$\Gamma_{\mu\nu}$. As discussed above, this would require the 
WI for the kernels ${\cal K}_{\mu\nu}$ and ${\cal K}_{\mu\nu\rho}$ 
appearing in them. 
Of course, in the context of the Abelian theory that we consider, the tree-level WIs ought to 
generalize naively to all orders, with no ghost contributions.
Thus, as a short-cut, one could simply derive the tree-level results and postulate
their validity to all orders. Instead we will derive the relevant 
all-order WIs formally, 
not only for completeness, but also 
in order to establish the necessary theoretical framework for addressing the 
same question in the more complicated case of non-Abelian theories.

The main subtlety involved in this treatment stems from the fact that 
the standard techniques furnish WIs for 
the connected kernels instead of the 1PI ones (in the sense described in the previous section), 
{\it i.e.}, for  ${\cal C}_{\mu\nu}$ instead of the 1PI ${\cal K}_{\mu\nu}$ appearing in the SDEs.
Therefore, in order to obtain the desired results, one must 
properly account for the 1PR terms, and 
subtract  their contributions from the WIs derived for the 
connected kernels ${\cal C}$.

In  order  to  determine  formally the all-order WI  satisfied  by  ${\cal C}$
kernels,  we  proceed  as  described  in~\cite{Itzykson:1980rh}.   
We    start   by   considering    the   Lagrangian   of
Eq.(\ref{Linv}); as  a consequence of  its invariance under  the gauge
transformations
\begin{equation}
\phi(x)\to e^{i\alpha(x)}\phi(x), \qquad \phi^\dagger(x)\to e^{-i\alpha(x)}\phi^\dagger(x), 
\qquad A_\mu\to A_\mu+\frac1{g}\partial_\mu\alpha(x),
\end{equation}
one has the conservation of the current    
\begin{equation}
J_\rho(x)=i:\phi^\dagger(x)\stackrel{\longleftrightarrow}{\partial_\rho}\phi(x):+2g:A_\rho(x)\phi^\dagger(x)\phi(x):, 
\qquad \partial^\rho J_\rho(x)=0.
\end{equation}
Then, one can derive WIs  relating Green's functions involving a single
current operator and  an arbitrary number of scalar  and photon fields,
as  a  result  of  current  conservation and  the  fact  that  Green's
functions are expressed as   time-ordered products in Minkowski space.  
Specifically, one has that
\begin{eqnarray}
& & \partial^\rho_x\langle0|TJ_\rho(x)\prod_{i=1}^nA_{\rho_i}(z_i)\prod_{j=1}^m\phi^\dagger(y_j)\phi(x_j)|0\rangle\nonumber\\
&=&\sum_{k=1}^n\langle0|T[J_0(x),A_{\rho_k}(z_k)]\delta(x^0-z^0_k)
\prod_{i=1,\,i\ne k}^nA_{\rho_i}(z_i)\prod_{j=1}^m\phi^\dagger(y_j)\phi(x_j)|0\rangle \nonumber \\
&+&\sum_{k=1}^m\langle0|T\prod_{i=1}^nA_{\rho_i}(z_i)\left\{[J_0(x),\phi^\dagger(y_k)]\delta(x^0-y^0_k)\phi(x_k) 
+\phi^\dagger(y_k)[J_0(x),\phi(x_k)]\delta(x^0-x^0_k) \right\}\times\nonumber\\ 
&\times&\prod_{j=1,\,j\ne k}^m\phi^\dagger(y_j)\phi(x_j)|0\rangle,
\end{eqnarray}
where the term containing $\partial^\rho J_\rho$ have been set to zero. On the other hand, 
canonical equal-time commutation relations (which ensure charge conservation) imply 
\begin{eqnarray}
\left[J_0(x),\phi(x')\right]\delta(x^0-x'^0)&=&g\phi(x)\delta^4(x-x'),\nonumber \\
\left[J_0(x),\phi^\dagger(x')\right]\delta(x^0-x'^0)&=&-g\phi^\dagger(x)\delta^4(x-x'),\nonumber \\
\left[J_0(x),A_\rho(x')\right]\delta(x^0-x'^0)&=&0.
\end{eqnarray}
We arrive then at the following general WI 
\begin{eqnarray}
& & \partial^\rho_x\langle0|TJ_\rho(x)\prod_{i=1}^nA_{\rho_i}(z_i)\prod_{j=1}^m\phi^\dagger(y_j)\phi(x_j)|0\rangle\nonumber\\
&=&-g\langle0|T\prod_{i=1}^nA_{\rho_i}(z_i)\prod_{j=1}^m\phi^\dagger(y_j)\phi(x_j)|0\rangle
\sum_{k=1}^m\left[\delta^4(x-y_k)-\delta^4(x-x_k)\right].
\label{IZ-WI}
\end{eqnarray}

This identity constitutes our starting point for deriving the WIs satisfied by the SD kernels ${\mathcal K}_{\mu\nu}$ and ${\mathcal K}_{\mu\nu\rho}$.

\begin{figure}[!t]
\includegraphics[width=14.5cm]{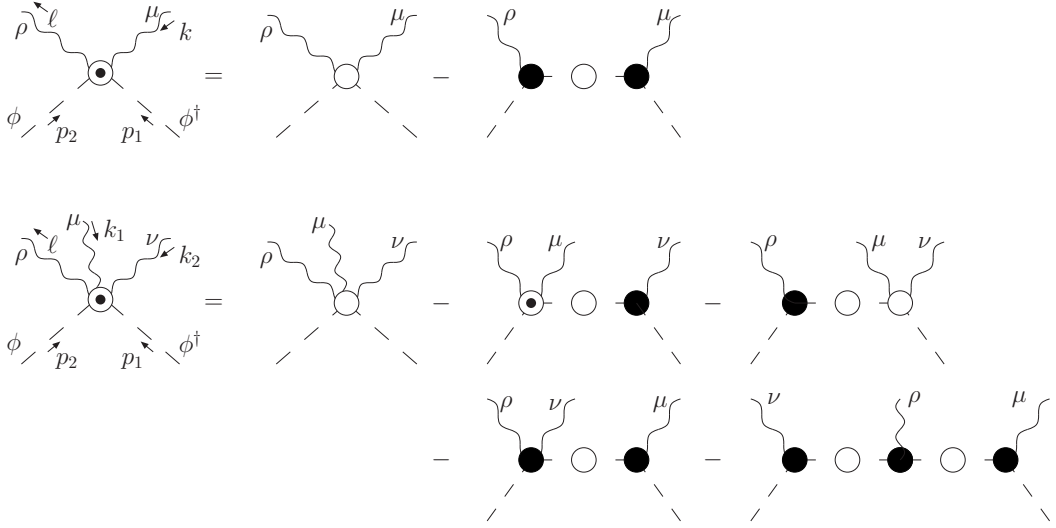}
\caption{\it The four- and five-particle kernels ${\cal K}_{\rho\mu}(k,p_1,p_2)$ and 
${\cal K}_{\rho\mu\nu}(k_1,k_2,p_1,p_2)$ appearing in the SDEs, 
and their relations with the corresponding amputated connected Green's function ${\cal C}_{\rho\mu}$ and ${\cal C}_{\rho\mu\nu}$ and 1PI Green's functions $\Gamma_\mu$ and $\Gamma_{\mu\nu}$.}
\label{SD-kernel}
\end{figure}

\subsection{Four particle kernel}
The first case of interest for us is when $n=2$ and $m=1$, {\it i.e.}, 
the photon--photon--scalar--scalar scattering kernel, defined as
\begin{eqnarray}
& & \int\!d^4x\,d^4x_1\,d^4y_1\,d^4z_1e^{i(p_1\cdot y_1+p_2\cdot x_1+k\cdot z_1-\ell\cdot x)} 
\langle0|TJ^\rho(x)A_{\beta}(z_1)\phi^\dagger(y_1)\phi(x_1)|0\rangle i\Delta^{(0)}_{\rho\alpha}(\ell)\nonumber \\
&=&(2\pi)^4\delta^4(k+p_1+p_2-\ell)\widetilde{\cal C}_{\alpha\beta}(k,p_1,p_2).
\end{eqnarray}
Contraction with $\ell^\alpha$ gives
\begin{eqnarray}
& & (2\pi)^4\delta^4(k+p_1+p_2-\ell)\ell^\alpha\widetilde{\cal C}_{\alpha\beta}(k,p_1,p_2)\nonumber \\
&=&-\frac{1}{\ell^2}\int\!d^4x\,d^4x_1\,d^4y_1\,d^4z_1e^{i(p_1\cdot y_1+p_2\cdot x_1+k\cdot z_1-\ell\cdot x)}\partial_\rho^x 
\langle0|TJ^\rho(x)A_{\beta}(z_1)\phi^\dagger(y_1)\phi(x_1)|0\rangle, 
\end{eqnarray}
and therefore, using Eq.(\ref{IZ-WI}) with $n=2$ and $m=1$, we get
\begin{eqnarray}
& & (2\pi)^4\delta^4(k+p_1+p_2-\ell)\ell^\mu\widetilde{\cal C}_{\alpha\beta}(k,p_1,p_2)\nonumber \\
&=&\frac{1}{\ell^2}\int\!d^4x_1\,d^4y_1\,d^4z_1\,e^{i[(p_1-\ell)\cdot y_1+p_2\cdot x_1+k\cdot z_1]}
g\langle0|TA_{\beta}(z_1)\phi^\dagger(y_1)\phi(x_1)|0\rangle\nonumber \\
&-&\frac{1}{\ell^2}\int\!d^4x_1\,d^4y_1\,d^4z_1\,e^{i[(p_2-\ell)\cdot x_1+p_1\cdot y_1+k\cdot z_1]}
g\langle0|TA_{\beta}(z_1)\phi^\dagger(y_1)\phi(x_1)|0\rangle.
\end{eqnarray}
Finally, defining the photon--scalar--scalar kernel $\widetilde{\cal C}_{\beta}$ as
\begin{eqnarray}
& & -i(2\pi)^4\delta^4(p_1+p_2-k)\widetilde{\cal C}_{\beta}(p_1,p_2)\nonumber\\
&=& \int\!d^4x\,d^4x_1\,d^4y_1\,e^{i(p_1\cdot y_1+p_2\cdot x_1-k\cdot x)}
\langle0|TA_\beta(x)\phi^\dagger(y_1)\phi(x_1)|0\rangle,
\end{eqnarray}
and introducing the amputated kernels
\begin{eqnarray}
& &\widetilde{\cal C}_{\alpha\beta}(k,p_1,p_2)=
i\Delta_{\alpha\rho}(\ell)i\Delta_{\beta\mu}(k) iS (p_1)
iS(p_2){\cal C}^{\rho\mu}(k,p_1,p_2),\nonumber \\
& & \widetilde{\cal C}_{\beta}(p_2,p_1)=i\Delta_{\beta\mu}(k) iS(p_1) iS (p_2)
\Gamma^{\mu}(p_2,p_1),
\end{eqnarray}
we obtain the required WI, 
\begin{equation}
\ell^\rho{\cal C}_{\rho\mu}(k,p_1,p_2) =
g \left[ S^{-1}(p_1)S(p_1-\ell)\Gamma_{\mu}(p_1-\ell,p_2)
- S^{-1}(p_2-\ell) S(p_2)\Gamma_{\mu}(p_1,p_2-\ell)\right],
\end{equation}
or, making use of momentum conservation,
\begin{equation}
\ell^\rho{\cal C}_{\rho\mu}(k,p_1,p_2) =
g \left[ S^{-1}(p_1)S(p_2+k)\Gamma_{\mu}(-k-p_2,p_2)
- S^{-1}(p_2)S(p_2-\ell)\Gamma_{\mu}(p_1,-k-p_1)\right].
\label{kWI1}
\end{equation}
Notice that, in our $U(1)$ case, one could equally well contract
with $\ell^\rho$  all the diagrams  appearing in the  decomposition of
${\cal   C}$   shown  in   Fig.\ref{SD-kernel},   using   the  WI   of
Eq.(\ref{WI3})   and   Eq.(\ref{WI4});  the   result   would  be  of course the
same. The contraction of Eq.(\ref{kWI1}) with $k^\mu$ can now be easily evaluated using the WI of Eq.(\ref{WI3}), and gives
\begin{equation}
k^\mu \ell^\rho{\cal C}_{\rho\mu}(k,p_1,p_2) =
g^2 \left\{ S^{-1}(p_1)+ S^{-1}(p_2)-S^{-1}(p_1) 
\left[S(p_2+k)+ S(p_1+k)\right]S^{-1}(p_2)  \right\}.
\label{WIC}
\end{equation}

As explained in the general analysis carried out in the previous section, we will need the WIs
satisfied by the kernel ${\cal K}_{\rho\mu}$, and not the ones for the connected Green's function ${\cal C}_{\rho\mu}$. These former 
WIs are however easily obtained, by making use of the relation (see Fig.\ref{SD-kernel})
\begin{equation}
i{\cal K}_{\rho\mu}(k,p_1,p_2)= i{\cal C}_{\rho\mu}(k,p_1,p_2)-i\Gamma_\rho(\ell-p_2,p_2)iS(p_2-\ell)i\Gamma_\mu(p_1,p_2-\ell).
\end{equation}
Contracting with $\ell^\rho$ and $k^\mu$, and using Eqs.(\ref{kWI1}), (\ref{WIC}) and (\ref{WI4}), 
we then arrive at the desired WIs, which read
\begin{eqnarray}
\ell^\rho{\cal K}_{\rho\mu}(k,p_1,p_2) &=& g\left[\Gamma_\mu(-k-p_2,p_2)S^{-1}(p_1)S(k+p_2)-\Gamma_\mu(p_1,-k-p_1)\right],
\label{kWIk}\\
k^\mu \ell^\mu{\cal K}_{\rho\mu}(k,p_1,p_2) &=&
g^2 \left[ S^{-1}(k+p_1)- S^{-1}(p_1)S(k+p_2)S^{-1}(p_2)\right].
\label{WIK}
\end{eqnarray}

\subsection{Five particle kernel}

The second case of interest for our construction is the one where $n=3$ and $m=1$, {\it i.e.}, 
the photon--photon--photon--scalar--scalar scattering kernel, defined as
\begin{eqnarray}
& & \int\!d^4x\,d^4x_1\,d^4y_1\,d^4z_1d^4z_2e^{i(p_1\cdot y_1+p_2\cdot x_1+k_1\cdot z_1+k_2\cdot z_2-\ell\cdot x)}\times\nonumber\\ 
&\times& \langle0|TJ^\rho(x)A_{\beta}(z_1)A_{\gamma}(z_2)\phi^\dagger(y_1)\phi(x_1)|0\rangle i\Delta^{(0)}_{\rho\alpha}(\ell)\nonumber \\
&=&i(2\pi)^4\delta^4(k_1+k_2+p_1+p_2-\ell)\widetilde{\cal C}_{\alpha\beta\gamma}(k_1,k_2,p_1,p_2).
\end{eqnarray}
Proceeding in exactly the same way as in the four-particle case, and introducing the amputated kernel
\begin{equation}
\widetilde{\cal C}_{\alpha\beta\gamma}(k_1,k_2,p_1,p_2)=
i\Delta_{\alpha\rho}(\ell)i\Delta_{\beta\mu}(k_1)i\Delta_{\gamma\nu}(k_2)
iS(p_1)iS(p_2){\cal C}^{\rho\mu\nu}(k_1,k_2,p_1,p_2),
\end{equation}
we obtain the WI
\begin{eqnarray}
\ell^\rho{\cal C}_{\rho\mu\nu}(k_1,k_2,p_1,p_2) &=&
g \left[ S^{-1}(p_1)S(p_1-\ell){\cal C}_{\mu\nu}(k_2,p_1-\ell,p_2)\right.\nonumber\\
&-&\left. S^{-1}(p_2-\ell) S(p_2){\cal C}_{\mu\nu}(k_2,p_1,p_2-\ell)\right].
\end{eqnarray}

Once again, the connected kernel ${\cal C}_{\rho\mu\nu}$ is not the one 
that appears in the SDEs, being related to the latter through the equation (see Fig.\ref{SD-kernel})
\begin{eqnarray}
i{\cal K}_{\rho\mu\nu}(k_1,k_2,p_1,p_2)&=&i{\cal C}_{\rho\mu\nu}(k_1,k_2,p_1,p_2)-i\Gamma_\rho(\ell-p_2,p_2)iS(p_2-\ell)i{\cal C}_{\mu\nu}(k_2,p_1,p_2-\ell)\nonumber \\
&-&i{\cal K}_{\rho\mu}(k_1,p_1+k_2,p_2)iS(p_1+k_2)i\Gamma_\nu(p_1,-p_1-k_2)\nonumber \\
&-&i\Gamma_{\rho\nu}(k_2,p_1+k_1,p_2)iS(p_1+k_1)i\Gamma_\mu(p_1,-p_1-k_1) \nonumber\\
&-&i\Gamma_\nu(-k_2-p_2,p_2)iS(p_2+k_2)i\Gamma_{\rho}(k_1+p_1,k_2+p_2)iS(p_1+k_1)\times\nonumber\\
&\times&i\Gamma_\mu(p_1,-p_1-k_1).
\end{eqnarray} 
Contracting with $\ell^\rho$, and making use of the WIs of Eqs.(\ref{kWI1}), (\ref{kWIk}), (\ref{WIC}) and (\ref{WI4}) we obtain, 
after a lengthy but straightforward calculation, the desired result
\begin{eqnarray}
\ell^\rho{\cal K}_{\rho\mu\nu}(k_1,k_2,p_1,p_2)&=&gS^{-1}(p_1)S(p_1-\ell){\cal C}_{\mu\nu}(k_2,p_1-\ell,p_2)-g \Gamma_{\mu\nu}(k_2,p_1,-k_2-k_1-p_1)\nonumber \\
&+&g\Gamma_\mu(-k_1-p_2,p_2)S(k_1+p_2)\Gamma_\nu(p_1,-p_1-k_2)\nonumber\\
&+&g\Gamma_\nu(-k_2-p_2,p_2)S(k_2+p_2)\Gamma_\mu(p_1,-p_1-k_1).
\label{WIK5}
\end{eqnarray} 

\section{\label{sde} PT Green's functions from Schwinger-Dyson equations}

In  this  section we will carry out in detail the PT construction at the level of the SDEs.
Specifically, from  the  SDEs shown  in
Figs.\ref{SDE-prop} and \ref{SDEs-vert}, we will derive the PT Green's functions for the scalar propagator
$S$, the trilinear vertex $\Gamma_{\mu}$, and the quadrilinear vertex $\Gamma_{\mu\nu}$.

\subsection{The scalar propagator}  

As far as the SDE of the scalar propagator is concerned, the first step will
be to  isolate all the  type A and type B diagrams, 
on which one could implement the characteristic PT
decomposition of Eqs.~(\ref{PTsplit1}) and (\ref{PTsplit3}).
\begin{figure}[!t]
\includegraphics[width=16cm]{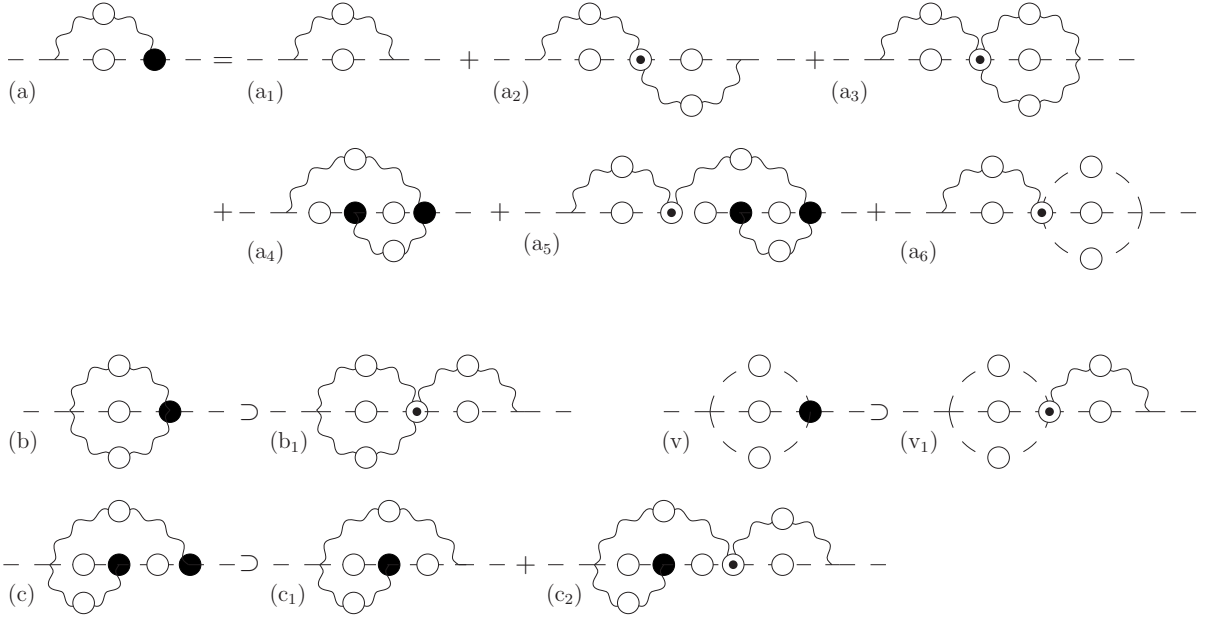}
\psframe*[linecolor=white](-8.5,4.6)(-8,4)
\caption{\it Isolating from the SDE of the scalar self-energy all the terms providing PT amplitudes.}
\label{diags-abc}
\end{figure}
To this end,  let us start with diagram  (a) of Fig.\ref{SDE-prop} and
unwrap the full $\Gamma_{\mu}$ by means  of its own SDE. The result is
shown in  the first two lines of  Fig.\ref{diags-abc}; clearly (a$_1$)
and (a$_2$) are  type A diagrams, while (a$_3$),  (a$_4$), (a$_5$) and
(a$_6$) are type B. There are  still four  diagrams  
of type B missing: the
first is obtained when unwrapping the full four-particle vertex of (b)
by substituting its SD series, retaining only the term where the
corresponding diagram  (i) of Fig.\ref{SDEs-vert}  appears; the second
diagram emerges 
when unwrapping the full four particle vertex of (v), keeping the term
in which diagram (n) of Fig.\ref{SDEs-vert} appears; finally, the remaining 
two diagrams of type B
come from diagram (c), after unwrapping the full trilinear vertex on
the right, retaining from the corresponding SDE  
the tree-level vertex and the diagram denoted by (d).  
All other  diagrams contributing to the scalar propagator
are of type C, {\it i.e.},  inert as far as
the  PT construction  is concerned, and will be  left untouched.
We emphasize that the above separation of diagrams in 
types A,B, and C is unique and unambiguous, regardless of 
the possibility that one has to further unwrap some of the 
full internal vertices, using 
their corresponding SDE. For example, the full internal
vertex $\Gamma_{\mu}$ appearing inside diagram ($a_4$) may be replaced
by the r.h.s. of the SDE in Fig.5, forcing the appearance of a bare (tree-level)  
scalar-scalar-photon vertex. 
This latter vertex is, however, internal,
i.e. all its legs are irrigated by virtual momenta; therefore, it is
not supposed to undergo the PT 
decomposition, and must remain as it is.
(For the same reason, in the two-loop construction of section \ref{ptpert}, the internal vertices in 
graphs (3a),(3c), and (3d) did not furnish any pinching momenta). 

At this point, one carries out on the above type A and B diagrams 
the usual PT decomposition 
given in Eqs.(\ref{PTsplit1}) and (\ref{PTsplit3}).
This will generate the following terms [we use hereafter the notation introduced previously in Eq.(\ref{gendef})]
\bea
(\mathrm{x})&=&(\mathrm{x})^{\mathrm{F}\mathrm{F}} + (\mathrm{x})^{\mathrm{P}0} 
+ (\mathrm{x})^{0\mathrm{P}}  - (\mathrm{x})^{\mathrm{P}\mathrm{P}} \qquad \mathrm{x}=\mathrm{a}_1, \mathrm{a}_2 \nonumber \\
(\mathrm{y})&=&(\mathrm{y})^{\mathrm{F}}+(\mathrm{y})^{\mathrm{P}} \qquad \mathrm{y}=\mathrm{a}_3, \mathrm{a}_4, \mathrm{a}_5, \mathrm{a}_6, \mathrm{b}_1, \mathrm{c}_1, \mathrm{c}_2, \mathrm{v}_1.
\eea

We will now analyze separately the terms appearing in the above equations.

\subsubsection{$\Gamma^\mathrm{P}\Gamma^{(0)}$ and $\Gamma^{(0)}\Gamma^\mathrm{P}$ terms}

\begin{figure}
\includegraphics[width=15cm]{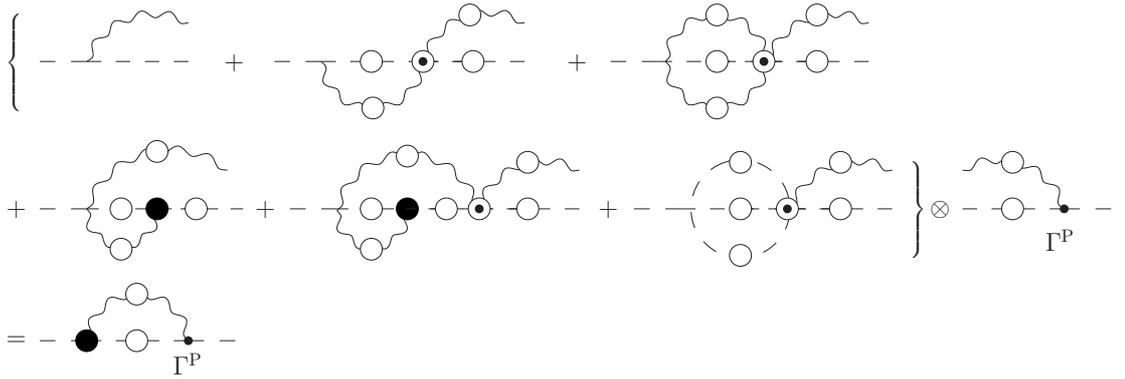}
\caption{\it The rearrangement of $\Gamma^\mathrm{P}\Gamma$ terms leading to a $\Gamma^\mathrm{P}$ acting on a full $\Gamma$ vertex.}
\label{ggP-rearrange}
\end{figure}

The first diagrams we will consider 
are those of type B. As already discussed, the strategy for treating these diagrams consists in   
factoring out the $\Gamma^{\mathrm P}$ vertex, and choosing the appropriate 
combination of graphs, in order to force the appearance of a full vertex $\Gamma$ on the opposite side of the diagram.  
The longitudinal momentum of the pinching vertex will act on this latter full vertex, 
thus triggering the corresponding WI.

In the case of the $\Gamma^\mathrm{P}\Gamma$ the combination one needs to consider is promptly found to be
$(\mathrm{a}_1)^{\mathrm{P}0}+(\mathrm{a}_2)^{\mathrm{P}0}+(\mathrm{a}_3)^{\mathrm{P}}+
(\mathrm{a}_4)^{\mathrm{P}}+(\mathrm{a}_5)^{\mathrm{P}}+(\mathrm{a}_6)^{\mathrm{P}}$;
it gives rise to the desired full vertex $\Gamma$, on which the $\Gamma^{\mathrm{P}}$ will act. 
Specifically,
\begin{eqnarray}
& &(\mathrm{a}_1)^{\mathrm{P}0} + (\mathrm{a}_2)^{\mathrm{P}0} +  (\mathrm{a}_3)^{\mathrm{P}} 
+ (\mathrm{a}_4)^{\mathrm{P}} + (\mathrm{a}_5)^{\mathrm{P}}+(\mathrm{a}_6)^{\mathrm{P}}\nonumber\\
&=&\int\![dk]\, i\Gamma^{\mathrm P}_{\mu}(-k-q,q)
i\Delta^{\mu\nu}(k)iS(k+q)i\Gamma_{\nu}(-q,k+q)\nonumber \\
&=&-g\int\![dk]\,\frac 1{k^2} S(k+q) k^\nu \Gamma_{\nu}(-k,-q)\nonumber \\
&=&-g^2\int\![dk]\,\frac {1}{k^2} S(k+q)
\left[S^{-1}(k+q)-S^{-1}(q)\right]\nonumber \\
&=& - i G(q)S^{-1}(q).
\end{eqnarray}
In obtaining the above expression we have used the WI of Eq.(\ref{WI3}), the SDE for the auxiliary function $G$ of Eq.(\ref{aux-function}), 
together with the dimensional regularization result of Eq.(\ref{dim_reg}).

For the symmetric term $\Gamma\Gamma^\mathrm{P}$  the combination one needs to
consider is different, and reads
$(\mathrm{a}_1)^{0\mathrm{P}}+(\mathrm{a}_2)^{0\mathrm{P}}+
(\mathrm{b}_1)^{\mathrm{P}}+(\mathrm{c}_1)^{\mathrm{P}}+(\mathrm{c}_2)^{\mathrm{P}}+(\mathrm{v}_1)^{\mathrm{P}}$, as shown in Fig.\ref{ggP-rearrange}; however,
the  result is  the same, and the latter combination will give rise  to the
mirror contribution  of the one just calculated. Therefore one has the result
\begin{eqnarray} 
& &(\mathrm{a}_1)^{\mathrm{P}0} + (\mathrm{a}_2)^{\mathrm{P}0} +  
(\mathrm{a}_3)^{\mathrm{P}} + (\mathrm{a}_4)^{\mathrm{P}}+(\mathrm{a}_5)^{\mathrm{P}}+(\mathrm{a}_6)^{\mathrm{P}}\nonumber \\
&+&(\mathrm{a}_1)^{0\mathrm{P}}+(\mathrm{a}_2)^{0\mathrm{P}}+
(\mathrm{b}_1)^{\mathrm{P}}+(\mathrm{c}_1)^{\mathrm{P}}+(\mathrm{c}_2)^{\mathrm{P}}+ (\mathrm{v}_1)^{\mathrm{P}}
=-2i G(q) S^{-1}(q).
\end{eqnarray}

\subsubsection{$\Gamma^{\mathrm P}\Gamma^{\mathrm P}$ terms}

Type A diagrams $(\mathrm{a}_1)^\mathrm{PP}$ and $(\mathrm{a}_2)^\mathrm{PP}$
are of central importance in our construction,  since, among other things, 
they
must generate {\it dynamically}  the BFM ghost sector, 
exactly as happened in the two-loop example of section \ref{ptpert}. 

Of these two diagrams, $(\mathrm{a}_1)^\mathrm{PP}$ gives simply
\begin{eqnarray}
(\mathrm{a}_1)^\mathrm{PP}&=&\int\![dk]\,i\Gamma^{\mathrm P}_{\mu}(-k-q,q)
i\Delta^{\mu\nu}(k) iS(k+q)i\Gamma^{\mathrm P}_{\nu}(-q,k+q)\nonumber \\
&=&-g^2\int\![dk]\, S(k)=
  \begin{picture}(58,27) (495,-130)
    \SetWidth{0.5}
    \SetColor{Black}
    \DashLine(514,-129)(524,-139){5}
    \DashLine(524,-139)(534,-129){5}
    \DashCArc(524,-120.86)(12.9,-39.16,219.16){5}
    \COval(524,-107)(4,4)(0){Black}{White}
    \DashLine(495,-139)(553,-139){5}
    \Vertex(524,-139){1.41}
  \end{picture}
\end{eqnarray} 
and thus generates an effective seagull-like contribution;  
it will be combined later with diagram (w) of Fig.\ref{SDE-prop}, in order to provide the graph $\widehat{\mathrm{w}}$.

\begin{figure}[!t]
\includegraphics[width=6.5cm]{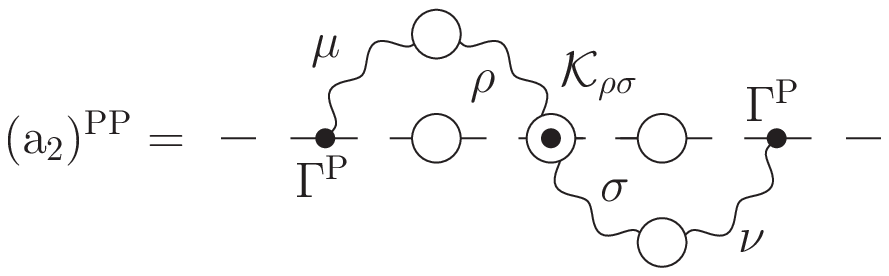}
\caption{\it Feynman diagram corresponding to the term $(\mathrm{a}_2)^\mathrm{PP}$.}
\label{a2PP}
\end{figure}

The second term, ($\mathrm{a}_2)^\mathrm{PP}$, is shown in Fig.\ref{a2PP}, and reads
\begin{eqnarray}
(\mathrm{a}_2)^\mathrm{PP} &=& \int\![d\ell]\int\![dk]\,i\Gamma^{\mathrm P}_{\mu}(-\ell-q,q)
i\Delta^{\mu\rho}(\ell)iS (\ell+q)
i{\cal K}_{\rho\sigma}(k,-k-q,\ell+q)\nonumber\\
&\times& iS(k+q)
i\Delta^{\sigma\nu}(k)i\Gamma^{\mathrm P}_{\nu}(-q,k+q)\nonumber \\
&=& -ig^2\int\! [d\ell]\int\![dk]\,\frac1{\ell^2}\frac1{k^2}
S(\ell+q)S(k+q)\ell^\rho k^\sigma{\cal K}_{\rho\sigma}(k,-k-q,\ell+q).
\label{a2PPeq}
\end{eqnarray}
We thus see how the SD kernel ${\cal K}_{\rho\sigma}$, studied 
in detail in Section \ref{ck}, makes its appearance.
Using the WI of Eq.(\ref{WIK}), and the SDE for the auxiliary function $G$ of Eq.(\ref{aux-function}), we find
\begin{eqnarray}
(\mathrm{a}_2)^\mathrm{PP}&=&
ig^4\int\![d\ell]\int\![dk]\,\frac1{\ell^2}\frac1{k^2} S(\ell+k+q)-ig^4 \int\![d\ell]\int\![dk]\,\frac1{\ell^2}\frac1{k^2}S(\ell+q)S(k+q)S^{-1}(q)\nonumber \\
&=&
  \begin{picture}(87,30) (295,-109)
    \SetWidth{0.5}
    \SetColor{Black}
    \DashCArc(338,-106)(21.02,267,627){2}
    \DashLine(300,-106)(377,-106){6}
    \COval(338,-106)(4,4)(0){Black}{White}
    \Text(315,-95)[]{\normalsize{\Black{$c$}}}
    \Text(315,-120)[]{\normalsize{\Black{$\bar c$}}}
  \end{picture}
+ iG^2(q)S^{-1}(q).
\label{PTstr}
\end{eqnarray}
where we have recognized that the first term  on the RHS of (\ref{PTstr})
is  exactly the one needed to generate the BFM ghost sector, whereas the second
contributes to the non-perturbative BQI of (\ref{BQIfinal}).
Actually, it is instructive to recognize that
this latter term is intimately connected with  
the perturbative rearrangement  of the 1PR  diagrams, referred to as ``strings'', 
i.e. all possible products of lower order self-energies, appearing when expanding perturbatively 
$\Delta_{\mu\nu}$ to a given order.  
In \cite{Binosi:2002vk,Binosi:2002ft} it has been shown that in QCD, the
terms one needs  to add to convert a string of order $n$ (in $g^2$)
containing more  than three self-energy insertions, into  a PT string,
will be canceled  by other strings of the same  order, but containing a
different  number   of  insertions;  the  only  case where this
cancellation will  not take place is  when the string  has exactly two
($\mathbb{S}_2^{(n)}$)  or  three  ($\mathbb{S}_3^{(n)}$)  self-energy
insertions.  Specializing these results  to the
case at hand, one has 
\begin{eqnarray}
\mathbb{S}_2^{(n)}&\to&\hat\mathbb{S}_2^{(n)}+2\sum_{i=1}^nG^{(n-i)}S^{-1,(i)}+\sum_{i=1}^nG^{(n-i)}S^{-1,(0)}G^{(i)}+4\sum_{i=2}^{n-1}\sum_{j=1}^{i-1}G^{(n-i)}S^{-1,(j)}G^{(i-j)}\nonumber \\
\mathbb{S}_3^{(n)}&\to&\hat\mathbb{S}_3^{(n)}-3\sum_{i=2}^{n-1}\sum_{j=1}^{i-1}G^{(n-i)}S^{-1,(j)}G^{(i-j)},
\end{eqnarray}
and therefore the contribution coming from the string conversion that must be added to the PT self-energy will be
\begin{equation}
\mathbb{S}^{(n)}=2\sum_{i=1}^nG^{(n-i)}S^{-1,(i)}+\sum_{i=1}^{n-1}\sum_{j=0}^{i-1}G^{(n-i)}S^{-1,(j)}G^{(i-j)},
\end{equation} 
or, to all orders,
\begin{equation}
\mathbb{S}=2GS^{-1}+G^{2}S^{-1}.
\end{equation} 
Of these two terms the first will cancel against 
an equal (but opposite in sign) contribution coming from the usual 
PT construction carried out on the 1PI diagrams, 
while the second one -- which, up to an immaterial $i$ 
factor is the second term of Eq.(\ref{PTstr}) -- represents the genuine string contribution.

\subsubsection{The $\Gamma^\mathrm{F}\Gamma^\mathrm{F}$ terms and the final rearrangement}

We finally consider the $\Gamma^\mathrm{F}\Gamma^\mathrm{F}$, {\it i.e.}, 
$(\mathrm{a}_1)^{\mathrm{F}\mathrm{F}},\ (\mathrm{a}_2)^{\mathrm{F}\mathrm{F}},
\ (\mathrm{a}_3)^{\mathrm{F}},\ (\mathrm{a}_4)^{\mathrm{F}},
\ (\mathrm{a}_5)^{\mathrm{F}},\ (\mathrm{a}_6)^{\mathrm{F}},$ $(\mathrm{b}_1)^{\mathrm{F}},\ (\mathrm{c}_1)^{\mathrm{F}}$, $(\mathrm{c}_2)^{\mathrm{F}}\, (\mathrm{v}_1)^{\mathrm{F}}$.

First of all,  notice that in the BFM the only Feynman rules 
different from the normal $R_\xi$  ones (excluding the ghost sector) are
those involving the $\widehat\phi A_\mu\phi^\dagger$ and 
$\widehat\phi \widehat\phi^\dagger\phi\phi^\dagger$
vertices; for them we have (see also the Appendix)
\begin{eqnarray}
& &i\Gamma_{\widehat\phi A_\mu\phi^\dagger}(q,-p_1)= 2ig p_{2\mu}=i\Gamma^\mathrm{F}(q,-p_1),\nonumber \\
& &i\Gamma_{\widehat\phi \widehat\phi^\dagger\phi\phi^\dagger}(k,p_1,p_2)=i(\lambda-g^2).
\end{eqnarray}  
Now, one should realize that our procedure has systematically replaced
all  of the  scalar-scalar-photon  vertices with  $\Gamma^\mathrm{F}$,
effectively  converting (as  far as  the Feynman rules describing their interactions are concerned) 
the  external legs, $\phi$ and  $\phi^\dagger$, 
into background ones,  $\widehat\phi$ and $\widehat\phi^\dagger$. Since
for  the  remaining  diagrams the  external  legs can  be  converted  into
background ones for free, we find
\begin{equation}
(\mathrm{a}_1)^{\mathrm{F}\mathrm{F}}+(\mathrm{a}_2)^{\mathrm{F}\mathrm{F}}+(\mathrm{a}_3)^{\mathrm{F}}+ 
(\mathrm{a}_4)^{\mathrm{F}}+(\mathrm{a}_5)^{\mathrm{F}}+(\mathrm{a}_6)^{\mathrm{F}}= (\widehat{\mathrm{a}}),
\end{equation} 
and
\begin{eqnarray}
& &(\mathrm{b}_1)^{\mathrm{F}}+\dots=(\widehat{\mathrm{b}}), \nonumber \\
& &(\mathrm{c}_1)^{\mathrm{F}}+(\mathrm{c}_2)^{\mathrm{F}}+\dots=(\widehat{\mathrm{c}}),\nonumber \\
& &(\mathrm{v}_1)^{\mathrm{F}}+\dots=(\widehat{\mathrm{v}}), 
\end{eqnarray}
where  the  ellipses denote all other  terms  appearing in  the
corresponding SDEs  ({\it i.e.}, the  missing terms on the RHS of the second and third line of Fig.\ref{diags-abc}).
Summarizing, the PT procedure has enforced the following identity 
[recall that the $\Gamma^\mathrm{P}\Gamma^\mathrm{P}$ terms appear with an extra minus sign, see Eq.(\ref{PTsplit3})]
\begin{eqnarray}
iS^{-1}(q)&=&-2i G(q)S^{-1}(q)-iG^2(q)S^{-1}(q)\nonumber \\
&+&g^2\int\![dk] S(k)+ig^4\int\![d\ell]\int\![dk]\frac1{\ell^2}\frac1{k^2}S(\ell+k+q)\nonumber \\
&+&(\widehat{\mathrm{a}})+(\widehat{\mathrm{b}})+
(\widehat{\mathrm{c}})+(\widehat{\mathrm{v}})+(\mathrm{w})+(\mathrm{z}).
\end{eqnarray}
The first two terms corresponds to the last two of the BQI of Eq.(\ref{BQIfinal}).
The third term can be added to the diagram (w) to get
\begin{equation}
(\mathrm{w})+g^2\int\![dk]\, S(k)=i(\lambda-g^2)\int\![dk]\, iS(k)\nonumber\\
=(\widehat{\mathrm{w}}).
\end{equation}
Moreover
\begin{equation}
ig^4\int\![d\ell]\int\![dk]\, \frac1{\ell^2}\frac1{k^2}S(\ell+k+q)
  \begin{picture}(77,26) (280,-110)
    \SetWidth{0.5}
    \SetColor{Black}
    \DashCArc(338,-106)(21.02,267,627){2}
    \DashLine(300,-106)(377,-106){6}
    \COval(338,-106)(4,4)(0){Black}{White}
    \Text(287,-107)[]{\normalsize{\Black{$=$}}}
    \Text(303,-117)[]{\small{\Black{$\widehat\phi$}}}
    \Text(377,-117)[]{\small{\Black{$\widehat\phi^\dagger$}}}
		\Text(328,-100)[]{\small{\Black{$\phi$}}}    
    \Text(320,-89)[]{\normalsize{\Black{$c$}}}
    \Text(320,-129)[]{\normalsize{\Black{$\bar c$}}}
    \Text(387,-104)[lc]{\normalsize{\Black{$=(\widehat{\mathrm{x}}),$}}}
  \end{picture}
\end{equation}
while, finally, $(\mathrm{z})\equiv(\widehat{\mathrm{z}})$.

Therefore, we find
\begin{eqnarray}
iS^{-1}(q)&=&-2i G(q)S^{-1}(q)- iG^2(q)S^{-1}(q)
+\bigg\{ (\widehat{\mathrm{a}})+(\widehat{\mathrm{b}})+
(\widehat{\mathrm{c}})+(\widehat{\mathrm{v}})+(\widehat{\mathrm{w}})+(\widehat{\mathrm{x}})+(\widehat{\mathrm{z}})\bigg\} \nonumber \\
&=&-2i G(q) S^{-1}(q)- iG^2(q) S^{-1}(q) + i\widehat{S}^{-1}(q),
\end{eqnarray}
which is exactly the BQI for the scalar self-energy.

\subsection{The trilinear vertex}

The construction of the PT vertices is in general significantly easier than the one carried out for the scalar propagator, 
due to the fact that there are no type A diagrams. In particular, for the case of the trilinear vertex $\Gamma_\mu$ only one type B diagram needs to be taken into account, namely diagram (d) of Fig.\ref{SDEs-vert}.
One has then
\begin{eqnarray}
({\mathrm d})^{\mathrm P}&=&\int\![d\ell]i\Gamma^{\mathrm P}_\nu(-q-\ell,q)i\Delta_{\nu\rho}(\ell)iS(q+\ell)i{\mathcal K}_{\rho\mu}(k,p_1,\ell+q)\nonumber\\
&=&-g\int\![d\ell]\frac1{\ell^2}S(q+\ell)\ell^\rho{\mathcal K}_{\rho\mu}(k,p_1,\ell+q)\nonumber\\
&=&g^2\int\![d\ell]\frac1{\ell^2}S(q+\ell)\Gamma_\mu(p_1,-k-p_1)\nonumber\\
&-&g^2\int\![d\ell]\frac1{\ell^2}S(\ell+q)S(k+\ell+q)\Gamma_\mu(-k-\ell-q,\ell+q)S^{-1}(p_1)\nonumber\\
&=&-iG(q)\Gamma_\mu(p_1,q)-iG_\mu(q,p_1)S^{-1}(p_1),
\end{eqnarray}
where we have used the WI of Eq.(\ref{kWIk}).
All other diagrams appearing in the SDE for the trilinear vertex are of type C, 
and remain unchanged; in that sense one can freely replace 
the external scalar leg with a background one, to get
\begin{eqnarray}
i\Gamma_\mu(p_1,q)&=&-iG(q)\Gamma_\mu(p_1,q)-iG_\mu(q,p_1)S^{-1}(p_1)+
\left\{({\mathrm d})^{{\mathrm F}}
+(\widehat{\mathrm{e}})+(\widehat{\mathrm{f}})+(\widehat{\mathrm{g}})+(\widehat{\mathrm{h}})\right\}\nonumber\\
&=&-iG(q)\Gamma_\mu(p_1,q)-iG_\mu(q,p_1)S^{-1}(p_1)+i\widehat\Gamma_\mu(p_1,q),
\end{eqnarray}
which is exactly the BQI of Eq.(\ref{BQIvert1}).

\subsection{The quadrilinear vertex} 

As the last step we will construct the PT quadrilinear vertex. Again, 
we have only one type B diagram that can give pinching contributions, namely diagram (i); 
its longitudinal momentum acting on the five particle kernel ${\mathcal K}_{\rho\mu\nu}$ will trigger the corresponding WI, Eq.(\ref{WIK5}). One has 
\begin{eqnarray}
({\mathrm i})^{{\mathrm P}}&=&\int\![d\ell]i\Gamma^{\mathrm P}_\nu(-q-\ell,q)i\Delta_{\nu\rho}(\ell)iS(q+\ell)i{\mathcal K}_{\rho\mu\nu}(k_1,k_2,p_1,\ell+q)\nonumber\\
&=&-g\int\![d\ell]\frac1{\ell^2}S(q+\ell)\ell^\rho{\mathcal K}_{\rho\mu\nu}(k,p_1,\ell+q)=\sum_{m=1}^4({\mathrm i})^{{\mathrm P}}_m,
\end{eqnarray}
Each $({\mathrm i})^{\mathrm{P}}_m$ represents a term in the WI of Eq.(\ref{WIK5}) which reads
\begin{eqnarray}
({\mathrm i})^{{\mathrm P}}_1&=&g^2\int\![d\ell]\frac1{\ell^2}S(\ell+q)\Gamma_{\mu\nu}(k_2,p_1,q)=-iG(q)\Gamma_{\mu\nu}(k_2,p_1,q)\nonumber \\
({\mathrm i})^{{\mathrm P}}_2&=&g^2\int\![d\ell]\frac1{\ell^2}S(\ell+q)S(p_1-\ell){\mathcal C}_{\mu\nu}(k_2,p_1-\ell,\ell+q)S^{-1}(p_1)\nonumber\\
&=&-iG_{\mu\nu}(k_2,\ell+q,p_1)S^{-1}(p_1)\nonumber \\
({\mathrm i})^{{\mathrm P}}_3&=&-g^2\int\![d\ell]\frac1{\ell^2}S(\ell+q)\Gamma_\mu(-k_1-\ell-q,\ell+q)S(k_1+\ell+q)\Gamma_{\nu}(p_1,-p_1-k_2)\nonumber \\
&=&-iG_\mu(q,k_2+p_1)\Gamma_\nu(p_1,-p_1-k_2)\nonumber\\
({\mathrm i})^{{\mathrm P}}_4&=&-g^2\int\![d\ell]\frac1{\ell^2}S(\ell+q)\Gamma_\nu(-k_2-\ell-q,\ell+q)S(k_2+\ell+q)\Gamma_{\mu}(p_1,-p_1-k_1)\nonumber \\
&=&-iG_\nu(q,k_1+p_1)\Gamma_\mu(p_1,-p_1-k_1).
\end{eqnarray}
For all the other (type C) diagrams appearing in the SDEs for the quadrilinear vertex, one can replace the external $\phi$ line with the corresponding background one $\widehat\phi$ without introducing new terms in the equation.
Using the results above, we finally arrive at the following equation 
\begin{eqnarray}
i\Gamma_{\mu\nu}(k_2,p_1,q)&=&-iG(q)\Gamma_{\mu\nu}(k_2,p_1,q)-iG_{\mu\nu}(k_2,\ell+q,p_1)S^{-1}(p_1)\nonumber \\
&-&iG_\mu(q,k_2+p_1)\Gamma_\nu(p_1,-p_1-k_2)-iG_\nu(q,k_1+p_1)\Gamma_\mu(p_1,-p_1-k_1)\nonumber\\
&+&({\mathrm i})^{{\mathrm F}}+(\widehat{\mathrm{j}})+(\widehat{\mathrm{k}})+(\widehat{\mathrm{l}})+(\widehat{\mathrm{m}})\nonumber \\
&=&-iG(q)\Gamma_{\mu\nu}(k_2,p_1,q)-iG_{\mu\nu}(k_2,\ell+q,p_1)S^{-1}(p_1)\nonumber \\
&-&iG_\mu(q,k_2+p_1)\Gamma_\nu(p_1,-p_1-k_2)-iG_\nu(q,k_1+p_1)\Gamma_\mu(p_1,-p_1-k_1)\nonumber\\
&+&i\widehat\Gamma_{\mu\nu}(k_2,p_1,q),
\end{eqnarray}
which is exactly the BQI satisfied by the quadrilinear vertex, Eq.(\ref{BQIvert2}).

\subsection{Renormalization issues}

So far we have succeeded in  converting the original SD series into an
equivalent one, where the external fields have  been substituted by
their background  counterparts. The procedure used has  been divided in
two  steps: ({\it  i}) carry  out the  PT algorithm  on the  (bare) SD
series and ({\it ii}) compare the result with the BQI satisfied by the
(bare)  Green's function  under  scrutiny. A question that rises naturally in this
context is whether this entire procedure is preserved by renormalization.

To answer this  question, one  should realize that  the BQIs are a direct
consequence of the  original BRST symmetry of the  theory;  therefore, 
within a suitable regularization scheme, such as dimensional regularization,
they will be  preserved by renormalization,  for the same reason  that the 
STIs do not get deformed.
Notice that  this  is  completely different  from  the case  of the 
Nielsen  identities \cite{Nielsen:1975fs}, describing the   gauge  fixing  parameter
dependence of the  bare Green's functions. In this latter case, one needs to
extend the BRST symmetry to  include the variation of the gauge fixing
parameter. This, in turn,  will spoil  the original BRST invariance  of the
theory,  implying   that  the   latter  identities  get   deformed  by
renormalization already at the one-loop level~\cite{Binosi:2005yk}.

To study an explicit example on how renormalization works for the BQIs, 
let us consider the renormalization of the two point function.
On the one hand we clearly have
\begin{eqnarray}
S^{-1}_ {\mathrm{R}}(g_{\mathrm{R}},\lambda_{\mathrm{R}},m^{2}_{\mathrm{R}};\mu) &=& Z_{\phi} S^{-1}(g,\lambda,m^2;\mu,\epsilon)\nonumber \\
\widehat S^{-1}_{\mathrm{R}}(g_{\mathrm{R}},\lambda_\mathrm{R},m^{2}_{\mathrm{R}};\mu) &=& 
Z_{\widehat\phi} \widehat S^{-1}(g,\lambda,m^2;\mu,\epsilon).
\label{S-renorm}
\end{eqnarray}
On the other hand, the function $\widetilde{G} \equiv1+G$ 
(where the 1 should be considered as its tree-level value, $\widetilde{G}^{(0)}\equiv1$) renormalizes multiplicatively as
\begin{equation}
\widetilde{G}_{\mathrm{R}}(g_{\mathrm{R}},\lambda_\mathrm{R},m^{2}_{ \mathrm{R}};\mu) = Z_{\widetilde{G}} \widetilde{G} (g,\lambda,m^2;\mu,\epsilon).
\label{G-renorm}
\end{equation}
Notice, however, that $Z_{\widetilde{G}}$ will not be an independent renormalization constant, 
because, due to the BQI of Eq.(\ref{BQIfinal}), 
its value is determined in terms of $Z_{\widehat\phi}$ and $Z_\phi$; specifically, 
\begin{equation}
Z_{\widetilde{G}} = Z_{\widehat\phi}^{\frac{1}{2}} Z_{\phi}^{-\frac{1}{2}}.
\end{equation}

To check the validity of this result at lowest order, we can 
carry out explicitly, at one-loop, 
the renormalization program for the relevant two-point functions. 

The one-loop expansion of Eqs.(\ref{S-renorm}) and (\ref{G-renorm}) reads
\begin{eqnarray}
\Sigma_{\mathrm{R}}^{(1)}(g_{\mathrm{R}},\lambda_{\mathrm{R}},m^{2}_{\mathrm{R}};\mu) &=& 
\Sigma_{\mathrm{R}}^{(1)}(g_{\mathrm{R}},\lambda_\mathrm{R},m^{2}_{\mathrm{R}};\mu,\epsilon)
+Z_\phi^{(1)}(q^2-m^2_{\mathrm{R}})-Z^{(1)}_{m^2}m^2_{\mathrm{R}}\nonumber \\
\widehat\Sigma_{\mathrm{R}}^{(1)}(g_{\mathrm{R}},\lambda_\mathrm{R},m^{2}_{\mathrm{R}};\mu) &=& 
\widehat\Sigma_{\mathrm{R}}^{(1)}(g_{\mathrm{R}},\lambda_\mathrm{R},m^{2}_{\mathrm{R}};\mu,\epsilon)
+Z_{\widehat\phi}^{(1)}(q^2-m^2_{\mathrm{R}})-\widehat Z^{(1)}_{m^2}m^2_{\mathrm{R}}\nonumber \\
{\widetilde{G}}^{(1)}_{\mathrm{R}}(g_{\mathrm{R}},\lambda_\mathrm{R},m^{2}_{\mathrm{R}};\mu) &=& \widetilde{G}^{(1)}(g,\lambda,m^2;\mu,\epsilon)+ Z_{\widetilde{G}},
\end{eqnarray}
while the one-loop divergent parts for the quantum 
and background two point functions, and the auxiliary function $G$, are given by
\begin{eqnarray}
\Sigma^{(1)}(g,\lambda,m^2;\mu,\epsilon)&=&-\frac4{\left(4\pi\right)^2\epsilon}g^2q^2-\frac2{\left(4\pi\right)^2\epsilon}m^2\left(\lambda+g^2\right)+\dots\nonumber\\
\widehat\Sigma^{(1)}(g,\lambda,m^2;\mu,\epsilon)&=&-\frac8{\left(4\pi\right)^2\epsilon}g^2q^2 
-\frac2{\left(4\pi\right)^2\epsilon}m^2\left(\lambda-g^2\right)+\dots\nonumber \\
\widetilde{G}^{(1)}(g,\lambda,m^2;\mu,\epsilon)&=&-\frac2{\left(4\pi\right)^2\epsilon}g^2+\dots,
\end{eqnarray}
where the ellipses denote finite parts. Combining these results we get
\begin{eqnarray}
& & Z^{(1)}_\phi=\frac1{4\pi^2\epsilon}, \qquad Z_{\widehat\phi}^{(1)}=\frac1{2\pi^2\epsilon}g^2, \nonumber \\  
& &Z^{(1)}_{m^2}\equiv \widehat Z^{(1)}_{m^2}=-\frac1{8\pi^2}\left(\lambda+3g^2\right), \qquad Z_{\widetilde{G}}^{(1)}=\frac1{8\pi^2\epsilon}g^2,
\end{eqnarray}
which shows that
\begin{equation}
Z_{\widetilde{G}}^{(1)}=\frac12\left(Z^{(1)}_{\widehat\phi}-Z^{(1)}_{\phi}\right),
\end{equation}
as expected. We end by noticing that the equality between the one-loop
mass renormalization  constants for quantum and  background scalar was to
be expected, at least due to two reasons: ({\it i}) the scalar and the PT (background) two-point
functions differ by the pinch contributions, 
which are all proportional to ($q^2 - m^2$); ({\it ii})
$G$ has engineering dimension zero, and thus  can give rise to
only one independent renormalization constant.

\section{\label{Conc} Discussion and Conclusions}

In this article we have presented  for the first time the extension of
the PT  at the level  of the SDE.   Specifically, we have  carried out
explicitly the PT procedure for the SDEs governing the dynamics of the
two- and three-point functions in scalar QED.  This Abelian theory has
non-trivial properties  under the pinching  action, due to  the simple
fact  that,  unlike normal  QED,  the  fundamental interaction  vertex
between a pair of charged scalars and a photon depends on the momentum
of  the  incoming  scalars.   This  in  turn  activates  the  pinching
procedure,  and gives  rise to  a  set of  modified effective  Green's
functions, which  coincide with the  BFM Green's function  computed in
the  Feynman  gauge,  to  all  orders  in  perturbation  theory.   The
extension  of this  procedure beyond  fixed-order  perturbation theory
requires certain operational adjustments, as discussed here in detail,
but does not introduce additional assumptions. The main result of this
paper is that the application of  the PT algorithm on the SDEs for the
conventional Green's functions in the usual covariant gauges generates
dynamically  the  SDEs  governing  the BFM  Green's  functions.   This
conversion of one  set of SDE to another  is highly non-trivial, given
that  the  Feynman rules  and  the  associated  ghost sector  is  very
different within these two gauge-fixing schemes.

As  has been  emphasized in  the Introduction,  the upshot  of  the PT
approach is to eventually  furnish a self-consistent truncation scheme
for the SDEs of gauge  theories.  It is therefore important to briefly
comment why the new SDEs obtained through pinching are superior to the
conventional  ones, and  how one  should proceed  to solve  them.  The
construction  carried out  here essentially  makes  manifest extensive
all-order rearrangements between the various terms in the SDEs, giving
rise   to  radically  different   structures.   The   ensuing  massive
cancellations are responsible for the special properties of the new PT
Green's  functions; instead,  in  the conventional  SDE expansion  the
consequences of  these rearrangments are obscured,  or even distorted,
by casual truncations  of the series.  In fact,  the advantages of the
new Schwinger-Dyson series can be  best exemplified in the case of QCD
itself.  Specifically, one of the most distinct features of the PT-BFM
scheme  is  the  special  way  in  which  the  transversality  of  the
background gluon self-energy is  realized. In particular, the study of
the  non-perturbative,  SD-type  of  equation  obeyed  by  the  latter
quantity reveals that, by virtue  of the Abelian-like WIs satisfied by
the vertices  involved, the transversality is  preserved {\it without}
the  inclusion  of  ghosts~\cite{Aguilar:2006gr}.  Thus,  gluonic  and
ghost    contributions   are    separately    transverse.    Moreover,
transversality  is enforced  without mixing  the orders  in  the usual
``dressed-loop''     expansion:    the     ``one-loop-dressed''    and
``two-loop-dressed''  sets of  diagrams are  independently transverse.
This is to  be contrasted with what happens  in the usual gauge-fixing
scheme of the covariant  renormalizable gauges, where the inclusion of
the ghost is  crucial for the transversality of  the gluon self-energy
already at  the level of  the one-loop perturbative  calculation.  The
importance of  this property in the  context of SDE is  that it allows
for a meaningful first approximation: instead of the system of coupled
equations involving gluon and ghost propagators, one may consider only
the  subset  containing   gluons,  without  compromising  the  crucial
property of  transversality.  Turning to the second  question, one may
proceed to  solve the new SDEs following  two, conceptually equivalent
but operationally  distinct, approaches.  For example, in  the case of
the  scalar propagator considered  in this  article, one  may continue
treating $S(q)$  as the unknown  dynamical variable, solve the  new SD
equation    in    terms   of    $S(q)$,    substitute   $S(q)$    into
(\ref{aux-function}) to obtain $G(q)$, and subsequently use the BQI of
(\ref{BQIfinal}) to construct  $\widehat S(q)$. Alternatively, one may
regard from the  beginning  $\widehat S(q)$  as  the new  dynamical
variable,  and use  (\ref{BQIfinal}) to  substitute everywhere  on the
r.h.s.   of  the  corresponding  SDE  $S(q)$  in  favor  of  $\widehat
S(q)$. It remains  to be seen which of these  two approaches will turn
out to be logistically more expeditious.

In  our opinion  the  most relevant  conceptual  contribution of  this
article is  the identification of  the precise procedure that  must be
followed   when   pinching   SDEs,   together   with   the   necessary
field-theoretic  ingredients that  one needs  to employ.   Despite the
fact  that we  have  restricted our  attention  to scalar  QED in  the
unbroken phase, the procedure described  should carry over, up to some
additional  book-keeping  complications to  the  broken  phase of  the
theory,  {\it i.e.}  when the  scalar field  develops  a non-vanishing
vacuum expectation  value, $v$,  endowing the photon  with a  mass and
adding a Higgs  scalar into the physical spectrum.   In that case, the
object of interest for the PT construction is the effective propagator
of  the  Higgs boson,  $\widehat\Delta_{H}$,  and  the possibility  of
constructing  the  non-perturbative  version  of the  (Abelian)  Higgs
effective     charge    $    v^2     \widehat\Delta_{H}$,    presented
in~\cite{Papavassiliou:1997fn}.   In addition,  and  more importantly,
the  present work  sets up  the stage  for the  generalization  of the
method in a non-Abelian context, and especially in QCD.

Turning to this important issue, we expect that, as far as the general
methodology is concerned,  the extension of this work  to the case QCD
should  go  through  with   no  additional  modifications.   From  the
technical  point  of view,  however,  one  needs  to overcome  several
obstacles.   In particular,  as  has become  obvious  by the  analysis
presented here, one needs to use  the result of the contraction of the
pinching momenta on the 1PI three-point functions and kernels.  In the
cases considered here,  the Abelian nature of the  theory gave rise to
simple  expressions  for the  WI  needed,  whose derivation,  although
laborious at  times, proceeded following textbook  techniques.  In the
case  of  QCD  the  object  of  central interest  will  be  the  gluon
self-energy;  the  upshot  of  the  PT construction  will  consist  in
transforming  its SDE  into  the corresponding  SDE  for a  background
gluon.  As  is known from  the perturbative all-order  construction of
the  gluon   self-energy,  one   of  the  necessary   ingredients  for
accomplishing this is  the STI for three-gluon vertex,  derived in the
classic work  by Ball and Chiu~\cite{Ball:1980ax}; in  the language of
our  Abelian  theory  this  STI   would  be  the  direct  analogue  of
Eq.(\ref{WI3}).  In addition, however,  one needs the STI satisfied by
the QCD  analogue of  the kernel ${\cal  C}_{\mu\nu}$, namely  the 1PI
kernel with four  off-shell gluons; to the best  of our knowledge this
result does not  exist in the literature.  Whereas  a derivation using
some    of   the    techniques    reviewed   here,    or   those    of
\cite{Pascual:1984zb}, may furnish the analogue of Eq.(\ref{kWI1}) for
the  four-gluon kernel, it  is not  clear whether  the result  will be
expressed  in  terms  of  quantities (e.g.,  auxiliary  ghost  Green's
functions)  that  could  be  directly  connected  to  those  appearing
typically  in the  PT construction.   This difficulty  may  be further
compounded by  the fact that in  QCD the ghost  sector is interacting,
and therefore  the auxiliary functions appearing  in the corresponding
BQI have  a much more  complicated structure than the  $G$, $G_{\mu}$,
and  $G_{\mu\nu}$, defined in  Eqs.(\ref{aux-function}), (\ref{aux1}),
and (\ref{aux2}),  respectively. Thus, in  the corresponding equations
instead  of   bare  ghost  propagators  and  vertices   we  will  have
fully-dressed ones.   In addition,  the simple WI  of Eq.(\ref{auxWI})
will be most certainly replaced by more involved expressions.  Despite
the  technical  complications mentioned  above,  we  believe that  the
extension of the  present work to QCD lies well  within our reach, and
hope to be able to present it in the near future.

\begin{acknowledgments}
D.B. gratefully acknowledges useful correspondence with A. Quadri. 
The research of J.P. was supported by Spanish MEC under the grant FPA 2005-01678. \\
Feynman diagrams have been drawn using \verb|JaxoDraw| \cite{Binosi:2003yf}.  
\end{acknowledgments}

\newpage

\appendix

\begin{figure}[!t]
\includegraphics[width=15cm]{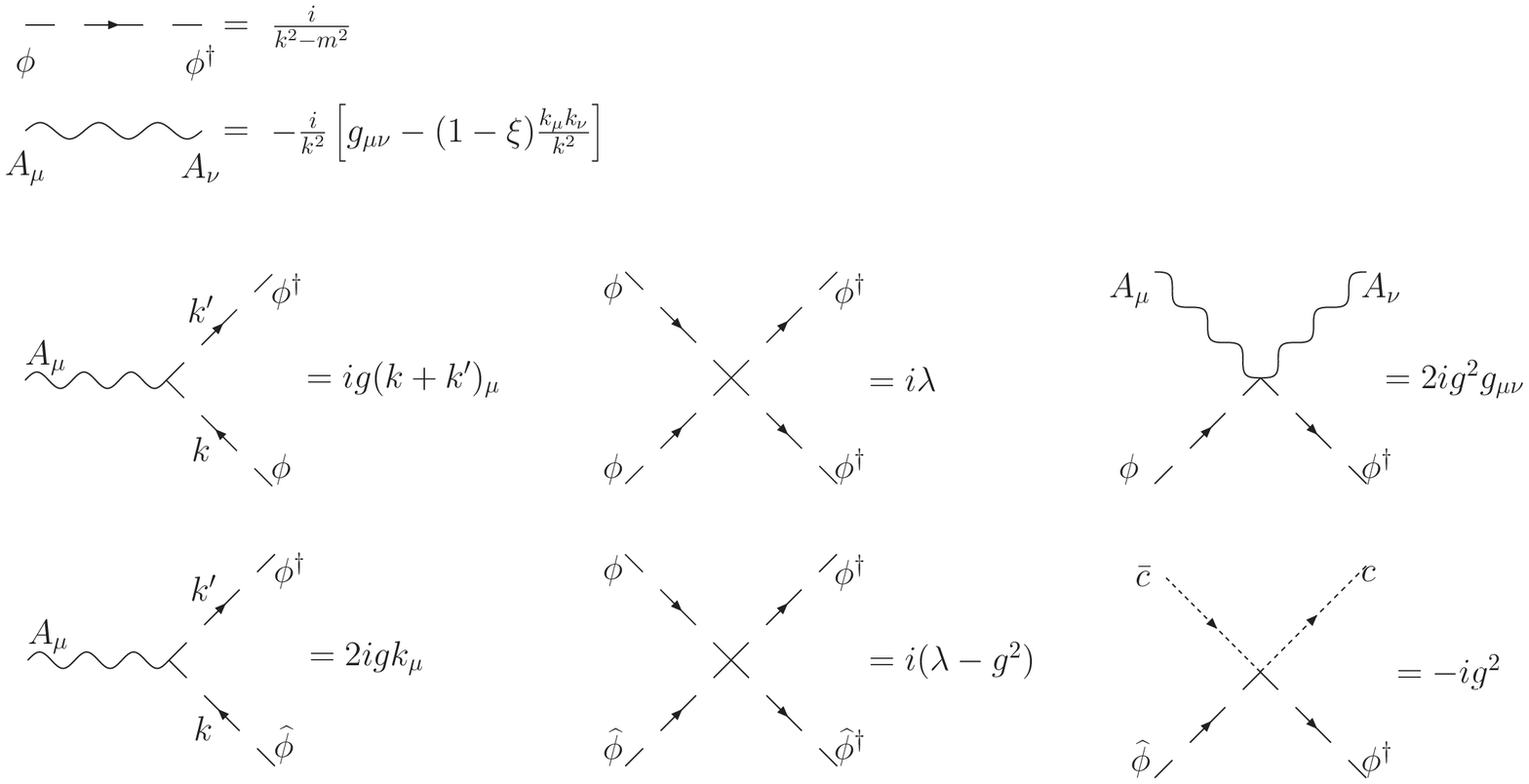}
\caption{\it Feynman rules for scalar QED used in the calculations both in the $R_\xi$ as well as the BFM Feynman gauge.}
\label{fr_sqed}
\end{figure}

\section{\label{frules} Feynman rules}

The Feynman rules for scalar QED (both in the $R_\xi$ and the BFM Feynman gauge) needed for the calculations carried out in the paper are listed in Fig.\ref{fr_sqed}. As already put forward in the paper, in order to obtain the full set of Feynman rules in the BFM gauge, 
one needs not only the gauge-fixing and Faddeev-Popov terms of Eq.(\ref{L-BFM}) but also the extra terms 
coming from the background-quantum splitting $\phi\to\widehat\phi+\phi$ carried out inside the 
gauge invariant Lagrangian (\ref{Linv}). The terms in which we are interested reads (in the Feynman gauge)
\be
\widehat{\cal L}\supset -2igA_\mu\left(\phi\partial^\mu\widehat{\phi}^\dagger+\phi^\dagger\partial^\mu\widehat\phi\right)+\left(\lambda-g^2\right)\widehat{\phi}^\dagger \widehat\phi\phi^\dagger\phi-g^2\bar cc\left(\widehat{\phi}^\dagger\phi+\widehat\phi^\dagger\phi\right),
\ee
and provides the Feynman rules shown above.

As far as the auxiliary functions are concerned the Feynman rules needed for their calculation can be obtained as follows.
\begin{enumerate}
\item For the BRST source terms they can be read directly from the BRST source lagrangian
\begin{equation}
{\cal L}_{\mathrm{BRST}}=\sum_\Phi\Phi^*s\Phi\supset A^*_\mu\partial^\mu c+ig\phi^{*\dagger}c\phi-ig\phi^*c\phi^\dagger.
\end{equation} 
Notice that the fact that $A^*_\mu$ has no interaction other than the one proportional to the derivative of its ghost field shown above, will enforce that the 1PI two point function $\Gamma_{cA^*_\mu}(q)$ is simply $-q_\mu$ to all orders.

\item For the background source terms one starts from the general identity
\begin{equation}
{\cal L}_{\mathrm{GF}}+{\cal L}_{\mathrm{FPG}}=s\Psi,
\end{equation}
with $\Psi$ the gauge fixing fermion 
\begin{equation}
\Psi=\bar c\left(\frac\xi2B+{\cal F}\right).
\end{equation}
Then if $\cal F$ is the background gauge fixing function of Eq.(\ref{BFMgff}) and we take into account the extended BRST transformations of Eq.(\ref{extBRST}) we get (in the Feynman gauge)
\begin{equation}
s\Psi\supset ig\bar c\Omega^{\phi^\dagger}\phi-ig\bar c\Omega^\phi\phi^\dagger
\end{equation}
\end{enumerate}

\begin{figure}[!t]
\includegraphics[width=16cm]{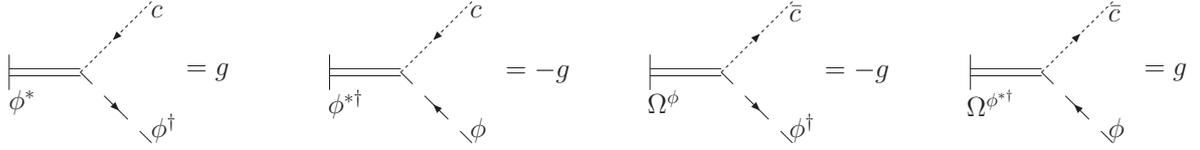}
\caption{\it Feynman rules for the interactions involving BRST and BFM sources.}
\label{fr_aux_sqed}
\end{figure}

The corresponding Feynman rules for both BRST and background field sources are given in Fig.\ref{fr_aux_sqed}.

\end{document}